# Prismatic Edge Dislocations in Graphite

James G. McHugh[1,2,3*], Pavlos Mouratidis[1], Anthony Impellizzeri[4], Kenny Jolley[1], Dogan Erbahar[4,5*], Chris P. Ewels[4]

[1]*Department of Chemistry, Loughborough University, Loughborough, LE11 3TU, UK*

[2]*Department of Physics and Astronomy, University of Manchester, Manchester, UK,*

[3]*National Graphene Institute, University of Manchester, Manchester, UK.*

[4]*Université de Nantes, CNRS, Institut des Matériaux Jean Rouxel, IMN, F-44000 Nantes, France*

[5]*Dogus University, Faculty of Engineering, Department of Mechanical Engineering, Ümraniye, 34775, İstanbul, Turkey*

## Abstract

Dislocations are a central concept in materials science, which dictate the plastic deformation and damage evolution in materials. Layered materials such as graphite admit two general types of interlayer dislocations: *basal* and *prismatic* dislocations, of which prismatic dislocations have been relatively less studied. Using density functional theory (DFT) calculations, we have examined different prismatic core structures in graphite and evaluated their structure, energetics and mobility. We find close energetic interplay between bonded and "free-standing" core structures in both zigzag and armchair directions, with a reconstructed stable zigzag core identified. We explore grain boundaries and prismatic dislocation pile-up, identifying metastable structures which may be important in energy storage. The role of interlayer stacking in core structure, dislocation glide and climb is also considered in-depth. Our calculations suggest that the prismatic dislocation core is stable up to high temperatures of approximately 1500K in bulk graphite. Above this temperature, the breaking of bonds in the dislocation core can facilitate climb, grain-boundary motion, and the annealing of damage through prismatic dislocation glide.

# 1. Introduction

Many of the most important properties of materials are determined not by the perfect bulk structure, but through the properties of defects such as dislocations. Dislocations represent one of the most fundamental concepts in materials science and, most importantly, they are the principle carriers of plastic deformation in crystalline materials [1-3]. Dislocations are one-dimensional line defects which accommodate strain through lattice deformation within a core region, and their dynamics largely accounts for the plastic deformation of materials under conditions of high temperature and under irradiation.

Graphite is an asymmetric, layered material, and bulk graphite consists of different stacking arrangements of individual graphene sheets. These layers differ from one another by an in-plane displacement along a C-C bond and can be labelled A, B and C depending on this relative shift. The asymmetry of graphite necessitates the existence of two types of interlayer dislocation: *basal* dislocations, where material is accommodated within a graphene sheet, creating regions of in-plane strain, and *prismatic* dislocations, where an additional partial graphene sheet is inserted into or removed from the bulk.

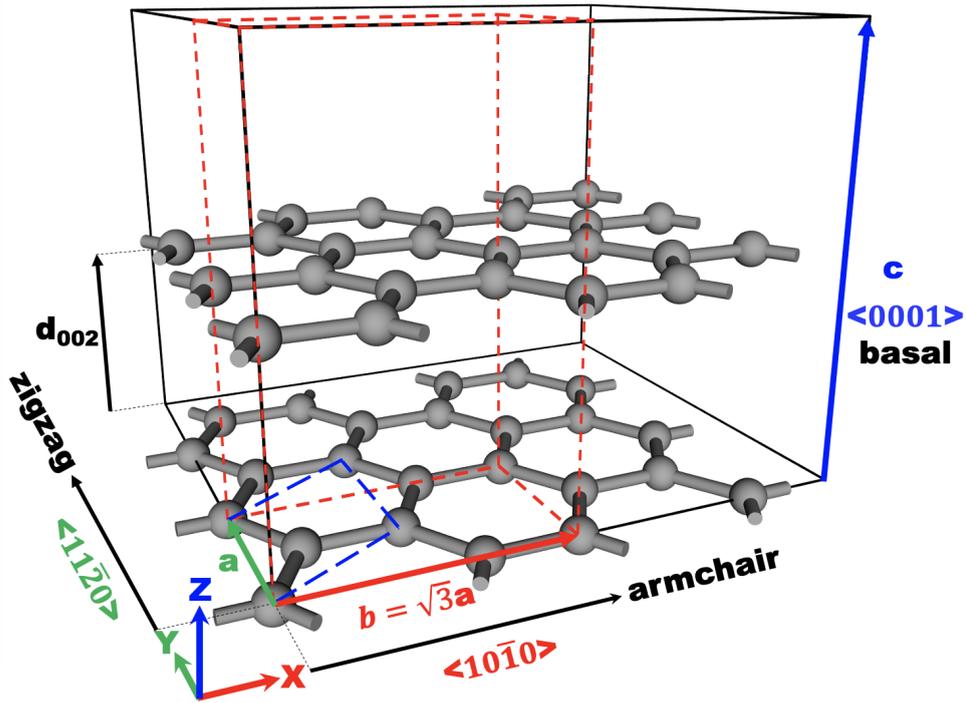

**Fig. 1:** Graphite lattice with different Bravais lattices (hexagonal in blue and orthorhombic in red), and high-symmetry directions indicated.

In Bernal (AB-stacked) graphite [4], the (0001) basal plane is closed-packed and the closed-packed directions are the $<11\bar{2}0>$ (or zigzag) directions, while the shortest lattice vectors are $1/3<11\bar{2}0>$, see **Fig. 1**. Perfect basal dislocation glide occurs within this plane with Burgers vector **b** = $1/3<11\bar{2}0>$. These perfect dislocations readily dissociate into the corresponding partials, which have been the subject of extensive experimental and theoretical investigation. [5-7] Prismatic edge dislocations have also been observed in graphite, using both STM [8,9] and TEM [10-13] imaging, under both normal and damaged conditions [14,15]. In carbon nanotubes their diffusion can mediate the formation of carbon nanotube walls [16]. They are readily formed under irradiation, where vacancies and interstitial atoms in excess of the equilibrium concentration can form prismatic dislocation loops [17,18]. Yet, despite their importance to many fundamental material properties, prismatic dislocations have received relatively little theoretical attention.

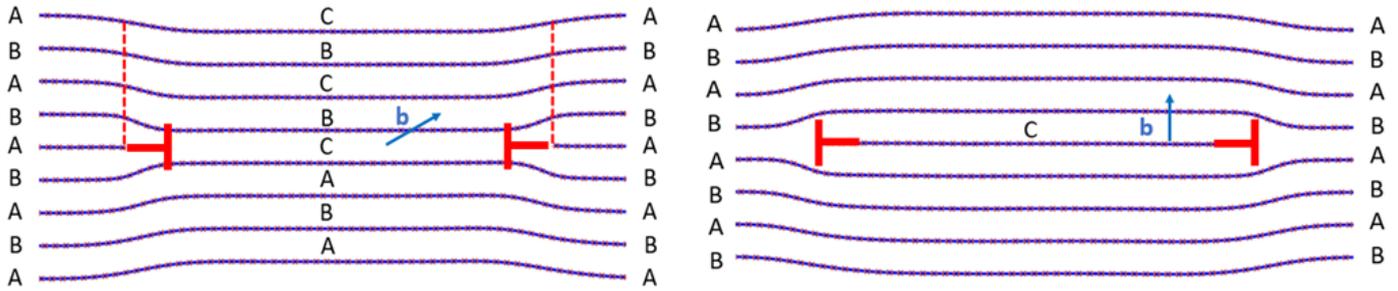

**Fig. 2:** Prismatic dislocations of **a)** vacancy character, which incorporates a basal component. This allows both "BC"-stacking across the dipole stacking fault and "AB"-stacking in adjacent sheets far from the core and **b)** interstitial character, where the prismatic sheet has perfect "C" character,.

**Fig. 2** schematically depicts a prismatic dipole configuration. A prismatic dislocation can be of either vacancy (**2a**) or interstitial (**2b**) character, where a graphene layer is correspondingly removed or inserted into graphite. The transition region between the extra or missing material and the perfect graphite lattice results in a pair of dislocation cores of opposite Burgers vector, which together comprise a dipole. The inserted layer of material experiences an energetic penalty, as there is a stacking fault between this layer and adjacent Bernal stacking layers. This is minimised when the additional layer has distinct (C) stacking to the adjacent A, B layers, leading to the low-energy rhombohedral ABC stacking fault [19,20]. While prismatic dislocations of interstitial character may automatically adopt this low-energy stacking, vacancy dislocations will initially result in unfavourable AA-stacking. This necessitates the in-plane shifting of material within a layer to restore preferable rhombohedral stacking, with the result that only layers adjacent to the stacking fault have distinct (AB, BC) stacking, which is restored to perfect Bernal lattice through passage of a basal dislocation, as shown in **Fig. 2a**.

The energetic and structural properties of different prismatic dislocation cores have been simulated in periodic supercells, using density functional theory (DFT) [21], where it was found that it is possible for interlayer bonds to form in the core region. In principle however, there are many possible bond reconstructions along an edge or between the adjacent sheets as seen, for example, in graphene nanotube edges. [22] In this work, we have undertaken a systematic study of the possible core configurations of the prismatic dislocation in graphite. In section 3.1, isolated prismatic cores are considered through explicit construction of graphene monolayer-bilayer junctions, which we find give a good approximation to the chemical structure of dislocation cores, while ignoring effects such as stacking & self strain. In this section, we focus particularly on the core reconstructions which are found to be most relevant in the bulk. In section 3.2, the structural and energetic properties of prismatic cores in the bulk are examined, both for isolated prismatic cores, and at regions of rotational disorder between graphite grains of different c-axis alignment, which correspond to different arrangements of dislocation arrays. In section 3.3, the dynamics of the bond-breaking process is considered, which is used to examine the stability of interlayer bonding, and dislocation motion through the mechanisms of dislocation glide and climb. Finally, in section 3.4 we compare simulated TEM images from our DFT structures to experiment.

## 2. Computational Methods

DFT calculations have been performed using the Quantum ESPRESSO ab-initio suite [23,24]. We have used the generalized gradient approximation (GGA) as parametrized by Perdew, Burke, and Ernzerholf (PBE) [25]. Interlayer interactions have been incorporated into our calculations through Grimme's D2 dispersion correction, which is found to give good agreement with experimental structural and energetic parameters [26-29]. Efficient Vanderbilt ultrasoft pseudopotentials are employed to approximate the effect of core electrons [30-32].

The wavefunction basis is expanded as a series of plane-waves with a maximum cut-off of $E_{cut}$ = 40 Ry and a charge density cut-off of $E_\rho$ = 500 Ry, which are found to be in close agreement to more highly-converged calculations. A non-zero electron temperature of kT = 0.02 eV for electronic level occupation is applied to aid convergence, using a Gaussian smearing function. The Brillouin zone is sampled using the Monkhorst-Pack scheme, with a $k$-point grid density of $2\pi \times 0.05 \text{ Å}^{-1}$. Structural optimisations were performed with a convergence threshold of 0.01 eV energy difference between subsequent iterations, and until the residual component of all forces is less than 0.01 eV/Å.

For cells containing vacuum, care has been taken to ensure that periodically-repeated images are well separated within the unit cell, and a separation of 20 Å is found to avoid any spurious self-interaction in the out of plane direction. All cells employed in our calculations employ an approximately equal length of material (~4 nm) in both monolayer and bilayer regions, so that we recover the interlayer separation of an isolated bilayer far from the cores. Reaction barriers are calculated using the climbing nudged elastic band (CI-NEB) method [33,34], where images are optimised so the force on each image orthogonal to the path reaches a value of less than 0.01 eV/Å.

High resolution transmission electron microscopy (HRTEM) images are simulated using the QSTEM software package [35] at room temperature on a 200 kV electron beam with spherical aberration -0.06 mm, defocus set to the Scherzer value of +15 nm, chromatic aberration 1 mm, and convergence angle 24.5 mrad.

Corresponding TEM images simulated for different defocus values are presented in Supplementary Materials.

## 3. Results
### 3.1 Isolated Cores

We first consider isolated prismatic dipoles, which are constructed as monolayer-bilayer transitions. This gives a pair of dislocations (a dislocation dipole) with opposite Burgers vector and parallel dislocation lines (see **Fig. 3**). Core structures therefore involve linear "cutting" of a graphene sheet in the dislocation line direction, resulting in a row of dangling bonds. These edges can be stabilised either *within* a layer by self-passivation through rehybridisation and restructuring of dangling carbon bonds (a 'free-standing edge' structure, FSE shown in **Fig. 3a**), or by *interlayer* bonding to the adjacent graphene layer (an 'inter-layer bonded' structure, ILB shown in **Fig. 3b**). This second option removes dangling bonds from the free edge, at the energetic cost of disrupting the local resonant π-bonded structure of the sheet below.

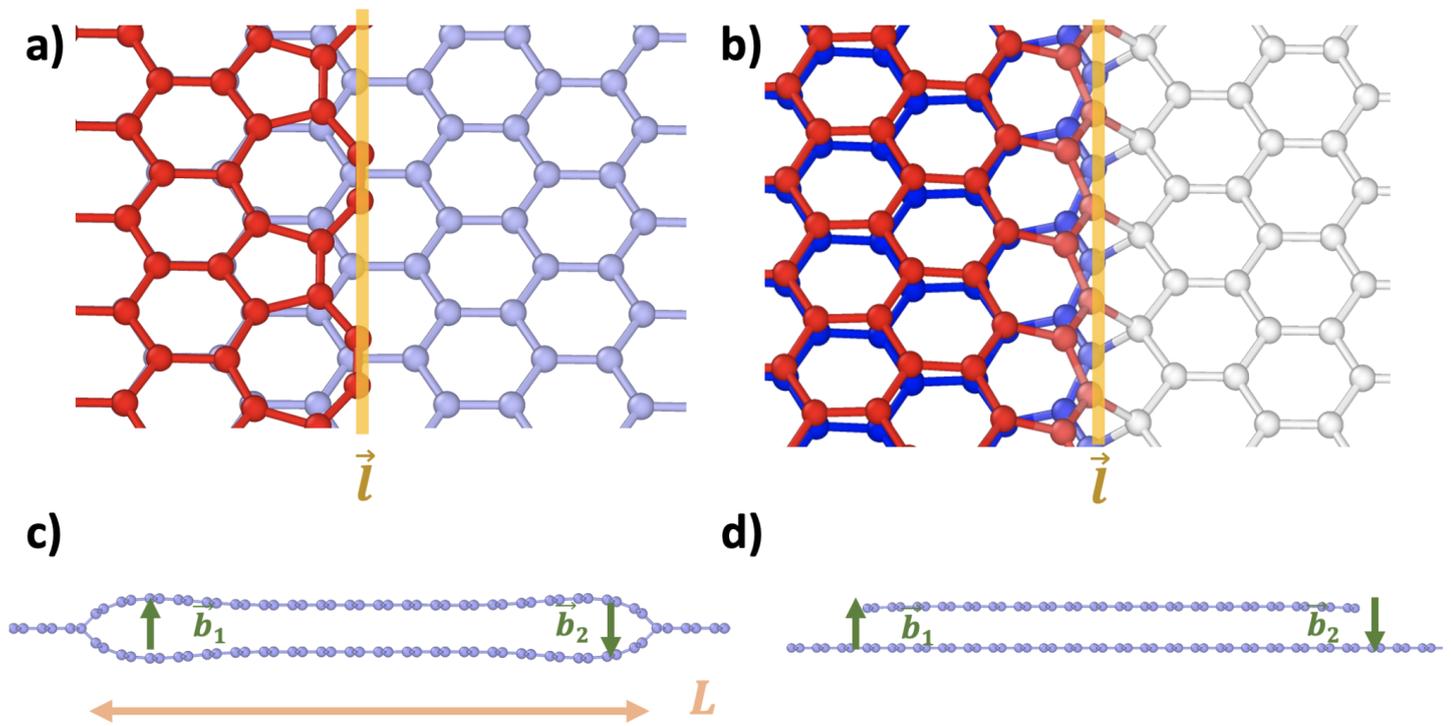

**Fig. 3: a)** Reconstructed free-standing edge (FSE) AA-stacked zigzag core. In the pictured cell, Dienes bond rotations have occurred along the top layer edge to reduce dangling carbon bonds. **b)** Interlayer-bonded (ILB) AA-stacked zigzag dislocation core. Dangling bonds in the top layer are eliminated by bonding into the layer below. The dislocation line direction, $\vec{l}$, which lies along the short direction in the ribbon geometry, is also indicated. Colours have been added for visual clarity, with the "ML", "Upper" and "Lower" sheets coloured white, red, and blue, respectively. **c)** and **d)** show side views of the same cells and the Burgers vector which characterises the extra material deposited at a core. Isolated dislocation cells have been constructed such that the net Burgers vector is zero, containing two dislocation cores of opposite sign.

In both cases, there are two possible contributions to the Burger's vector: a c-axis component, which is automatically included through the transition from monolayer to bilayer, and an in-plane component which is parallel to the basal plane. For ILB cores, the latter may be included by varying the chemical structure within the core, while a simple translation is sufficient in the case of FSE cores. The relative stability of the different possible core structures is therefore not a-priori obvious and will depend on the local chemical bond structure in the dislocation core region. This approach allows evaluation of different edge configurations, while ignoring secondary effects such as stacking faults and elastic energy.

The formation energy of a prismatic dislocation core, $E_f$, is given by the equation

$$E_f = (E_{cell} - n_{ml}\epsilon_{ml} - n_{bl}\epsilon_l)/2, \quad (1)$$

where $E_{cell}$ is the total energy of the cell containing both defects, $n_{ml}$ and $n_{bl}$ are the number of carbon atoms in the monolayer and bilayer regions respectively $\epsilon_{ml}$ and $\epsilon_{bl}$ are the calculated energy per carbon atom of isolated infinite monolayer and bilayer sheets, and we have divided by 2 due our use of dipole cells. The dislocation line energy, which is the formation energy of a dislocation core per unit length of dislocation, has also been calculated. This quantity determines the dislocation line tension which acts in the direction of the dislocation line and tends to shorten it, thus determining the energetic stability of a straight dislocation. [36,37]

### 3.1.1 Free-standing edges

Before considering the effect of interlayer-bonding we first summarise the properties of monolayer FSE graphene edges. Graphene edges can lie along either the zigzag <11$\bar{2}$0> or armchair <10$\bar{1}$0> directions, or a mixture of the two. **Fig. 4** shows the edge terminations and associated notation used in this work. The free-standing zigzag (FSE ZZ) edge (**Fig. 4a**) consists of a line of energetically costly dangling bonds. It is highly unstable and a number of reconstructions are possible [38], which depend significantly on local hydrogen content [39,40]. At room temperature, for low hydrogen content, or under electron irradiation [38, 41] a bond rotation with a barrier of 0.4-0.7eV occurs [38,40,41], giving the more stable Dienes edge (FSE 57) (**3b**). This edge, constructed from pentagons and heptagons, is stabilised by distorted rehybridisation to *sp*-triple bonding in the edge $C_2$ pairs. Both of these edges lie along the glide plane, however cleavage is also possible along the parallel shuffle plane (see **Fig. 4a**). The latter gives the highly unstable Klein edge, with individual single neighbor carbon atoms which can stabilize through pentagonal pairwise reconstruction of adjacent dangling bonds, resulting in the reconstructed (FSE RK) edge (**3c**).

Armchair edges are shown in **Fig. 4d-g**. The armchair edge spontaneously reconstructs within a graphene layer by rehybridising the external $C_2$ pair into a distorted ~1.24 Å triple bond [19, 20]. Since the armchair shuffle and glide planes have identical crystallographic character, their energetic and structural properties are the same (**Fig. 4d,e**). We include here two other armchair edges, the AC-56 edge [38], where removal or adsorption of a carbon atom at the edge carbon dimer results in a pentagonal reconstruction (**Fig. 4f**), and the AC-677 edge, which features a bond rotation similar to the zigzag FSE 57 edge (**Fig. 4g**). [38]

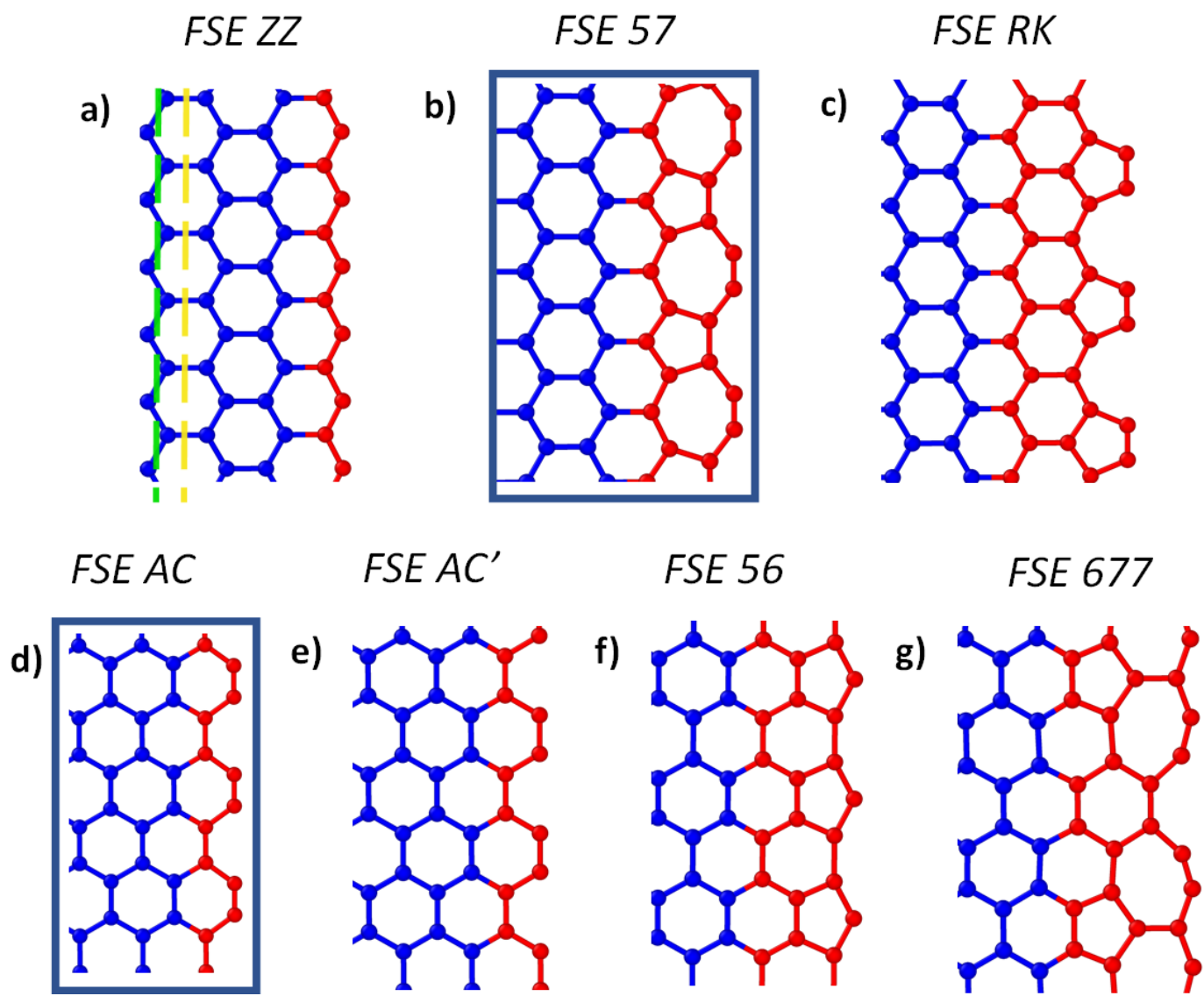

**Fig. 4:** Non-functionalised terminations of graphene free-standing edges (FSE). Edge atoms are marked in red. Lowest energy reconstruction along both crystallographic orientations have been highlighted.. **a)** Zigzag FSE-ZZ, with glide and shuffle planes are indicated by dotted yellow and green lines, respectively **b)** Dienes reconstructed FSE-57 **c)** Reconstructed Klein FSE-RK **d)** Armchair edge FSE-AC **e)** Shifted armchair edge FSE-AC' **f)** FSE-56 edge, resulting from gain or loss of carbon atoms **g)** FSE-677 edge from a bond rotation along the armchair edge.

We have performed structural relaxation of FSE structures with and without an underlying graphene substrate, the latter providing a baseline for comparison with the interlayer-bonded cores. Graphene edges are modelled using a finite-width nanoribbon, which is suspended over a perfect graphene sheet in the bilayer case, as shown in **Fig 3b,d**. **Table 1** shows the edge formation energies, which have been calculated using Eq. (1), and the effect of interlayer stacking has been accounted for by using appropriate values of the bilayer energy $\epsilon_{bl} = \epsilon_{AA}/\epsilon_{AB}$.

Our results for isolated edges confirm previously reported well-known trends [38,43,44]. The most stable edge structure in the $<10\bar{1}0>$ direction is the unreconstructed armchair edge (AC FSE). While the FSE ZZ edge is ~0.3 eV/Å more energetically costly than this, reconstruction into the FSE 57 edge lowers this energy to ~0.07 eV/Å more stable than the AC FSE edge, again in good agreement with previous work [38,43,44]. All structural properties such as bond lengths, and the energy differences between core configurations remain very similar in the presence or absence of a graphene substrate, regardless of the stacking configuration in the bilayer region.

| | $<11\bar{2}0>$ ZigZag Line energy (eV/Å) | | | |
|---|---|---|---|---|
| | Isolated | AA | AB | $\Delta_{AA-AB}$ |
| FSE ZZ | 1.357 | 1.317 | 1.319 | -0.002 |
| FSE RK | 1.379 | 1.353 | 1.363 | -0.009 |
| FSE 57 | 0.991 | 0.993 | 1.002 | -0.009 |
| | $<10\bar{1}0>$ Armchair Line energy (eV/Å) | | | |
| | Isolated | AA | AB | $\Delta_{AA-AB}$ |
| AC FSE | 1.064 | 1.112 | 1.038 | 0.073 |
| AC 56 | 1.506 | 1.559 | 1.491 | 0.068 |
| AC 677 | 1.155 | 1.240 | 1.181 | 0.026 |

**Table 1** Free-standing edge line energies for different reconstructions in finite-width graphene nanoribbon geometries. We include results for isolated nanoribbon edges, and nanoribbons deposited on a graphene substrate in AA and AB stackings (broken bond prismatic dislocation configurations)

### 3.1.2 Interlayer-bonded Zigzag Cores

ILB dislocation cores can be considered as the convergence of three graphene edges at the same line. Here, ILB dislocation cores have been constructed using graphene edges which lie along different glide planes. This alters the chemical structure with the core, while allowing the incorporation of different in-plane Burger's vector components. In the zigzag case, for example, including a single FSE RK edge in the core will shift the bilayer to AB stacking in comparison to a dislocation core composed of three FSE ZZ edges, shown in **Fig. 5**.

Core cells have been labelled according to the three edge configurations within the core region, terminating the monolayer "ML" (white atoms in **Fig. 5**), "Top" and "Bottom" bilayer graphene sheets (red, blue atoms in **Fig. 5**), which we will refer to by the convention ML-Top-Bottom. **Table 2** details the formation energies of the resulting ILB dislocation cores. For clarity, we also show the energy of the most favourable FSE configuration, and the difference between this and each ILB core. For all core configurations, the interlayer bonds bridging the two graphene sheets have sp$^3$ character due to their four-fold coordination. These sp$^3$ interlayer bonds are generally dilated compared to diamond, 1.53 Å or larger, due to additional strain in the core region.

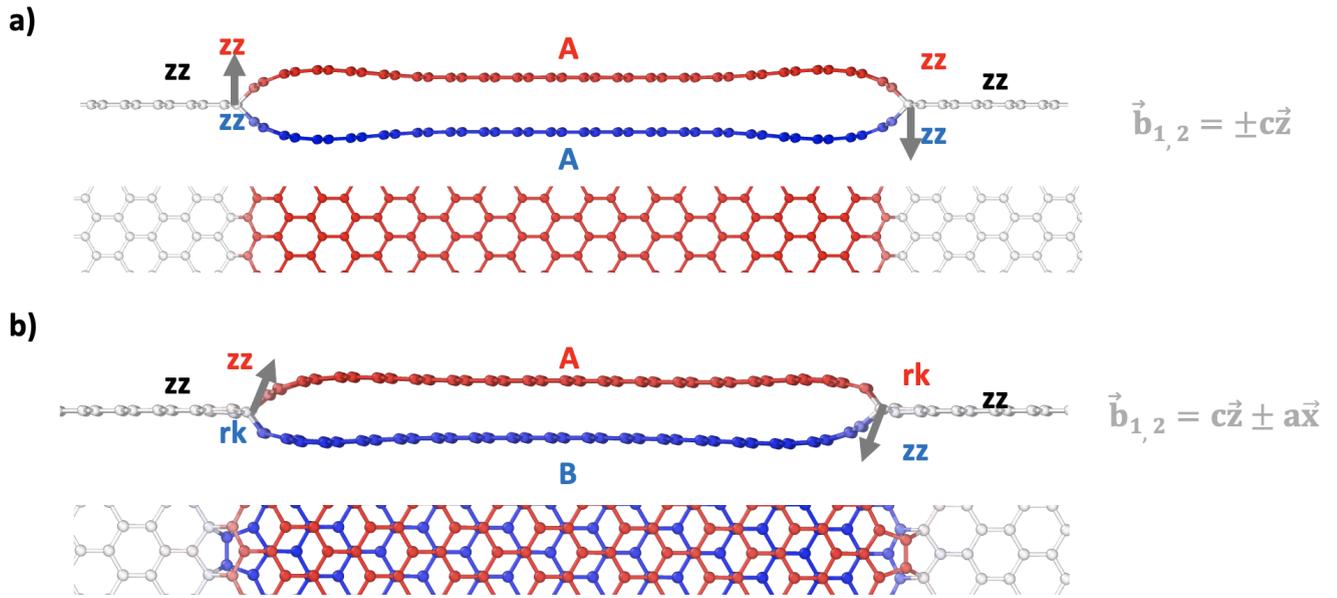

**Fig. 5:** Link between prismatic dislocation core edge configuration and Burgers vector. Core structures with different Burgers vector have been constructed through different combinations of graphene edges. This allows distinct stacking across the dislocation line and bonding within the core. **a)** A core constructed from the intersection of three ZZ edges leads to AA stacking in the bilayer region. **b)** The RK edge is shifted by a bond length with respect to a ZZ edge. This gives both a different core structure and AB stacking in the bilayer region.

We will largely confine our discussion to those structures which preserve the hexagonal graphene lattice within the core. Two ILB configurations are energetically favourable or comparable to the FSE 57 edge (the most stable free-standing edge configuration), both of which lead to AA-stacking. The symmetric ZZ-ZZ-ZZ configuration (**Fig. 6a**) consists entirely of zigzag edges, giving perfect hexagonal graphene that is smoothly connected and shared between the monolayer and bilayer regions. While this core is slightly higher in energy (0.03 eV/Å) than the reconstructed FSE 57 edge, the energy difference is very small. For example, there is a similar energy difference between the FSE 57 and FSE AC edges, both of which are observed in TEM images [41], and it is likely that some mixture of these two structures can co-exist within graphite. The ZZ-ZZ-ZZ edge structure can also bond into atoms on the opposite (β) sublattice rather than the same (α) sublattice by translating the bottom layer by a bond length, as shown in **Fig. 6b**. This gives AB stacking across the fault at the cost of a large increase in core formation energy. Henceforth, we will refer to these two configurations as the ZZ-ZZ-ZZ-α and ZZ-ZZ-ZZ-β structures, respectively.

| | | | | *Zigzag cores* | |
|---|---|---|---|---|---|
| ML | Top | Bottom | Stacking | Line energy (eV/Å) | Δ (eV/Å) |
| | FSE 57 | | AA | 0.993 | 0.000 |
| RK | ZZ | ZZ | AA | 0.976 | -0.017 |
| ZZ | ZZ | ZZ | AA | 1.025 | 0.032 |
| RK | RK | RK | AA | 1.526 | 0.533 |
| ZZ | RK | RK | AA | 1.693 | 0.700 |
| RK | ZZ | RK | AB | 1.245 | 0.252 |
| ZZ | ZZ | RK | AB | 1.342 | 0.349 |
| ZZ | ZZ | ZZ | AB | 1.603* | 0.032 |
| | | | *Armchair cores* | | |
| ML | Top | Bottom | Stacking | Line energy (eV/A) | Δ (eV/A) |
| | AC FSE | | AB | 1.038 | 0.000 |
| AC | AC | AC | AA | 0.933 | -0.105 |
| AC | AC | AC' | AB | 1.132 | 0.094 |

\* This is the energy of the ZZ-ZZ-ZZ core in a β-configuration (interlayer bonded atoms lie on different sublattices)

**Table 2:** Formation line energy for different interlayer-bonded prismatic dislocation cores, for both the zigzag and armchair cores. Cores which preserve aromaticity are highlighted in green. The last column gives the energy difference compared to the lowest energy free-standing configuration, FSE 57 and AC FSE, for the zigzag and armchair cases, respectively. Negative values imply stability of the ILB core relative to the free-standing edge. Selected structures are shown in Fig. 6 & Fig. 7, and additional structures are shown in Fig. S1.

It is interesting that the RK-ZZ-ZZ configuration, where a monolayer reconstructed Klein edge bonds into bilayer zigzag edges in the core region, (**Fig. 7a**) has the lowest energy of all FSE and ILB zigzag structures considered in this work (i.e. a RK edge binding into the basal plane below it). The resulting structure gives a line of shared 5-8 defects at the boundary. Every other zigzag-oriented core has a Burgers vector which lies only in the x-z plane and therefore contains only edge components. However the RK-ZZ-ZZ is the only zigzag core we have considered which also includes a screw component (i.e. movement along the dislocation line), in the lowest-energy configuration. In the absence of a screw component there is a triangular arrangement in the core (see inset of **Fig. 7a**) which is sheared away upon full relaxation.

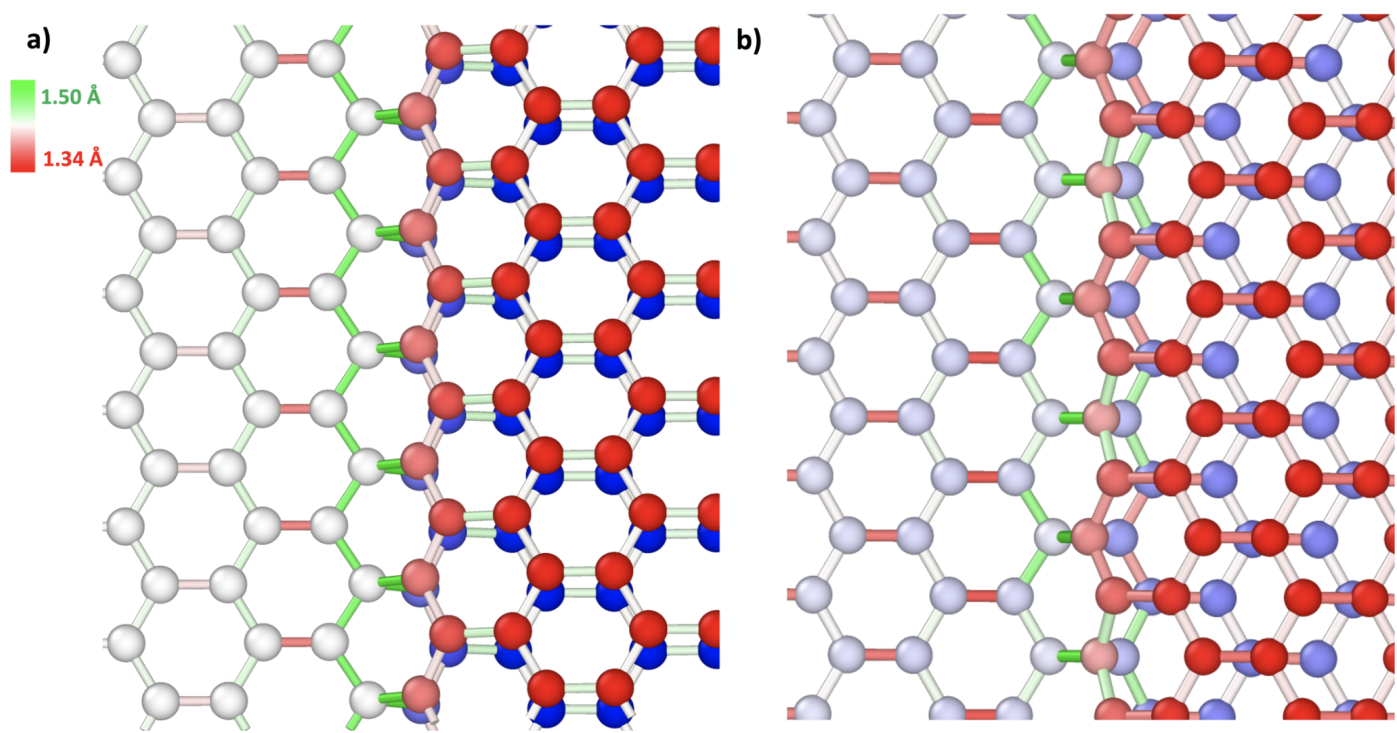

**Fig. 6: a)** ZZ-ZZ-ZZ-α core, where ML and both BL regions have zigzag edges, and **b)** ZZ-ZZ-ZZ-β core, which again has the same edge structure but bonds into the bottom layer along the opposite sublattice through a relative displacement of one bond length. C-C bonds are coloured depending on the degree of bond strain as shown in the key on the left, with white matching the pristine graphene bond length (1.42 Å).

All of the other cores involve multiple RK edges meeting in the core region, as well as more energetically costly pentagonal rings, and are generally much higher in energy than the FSE 57 structure. An example of this is given by the RK-ZZ-RK core (**Fig. 7b**), which is the most stable AB-stacked core, and features a high density of pentagonal and heptagonal defects, as well as extensive local strain accommodation as indicated by the green bonds leading to a relatively large line energy. Some further images of these cores are available in the Supplementary Material, **Fig. S1**.

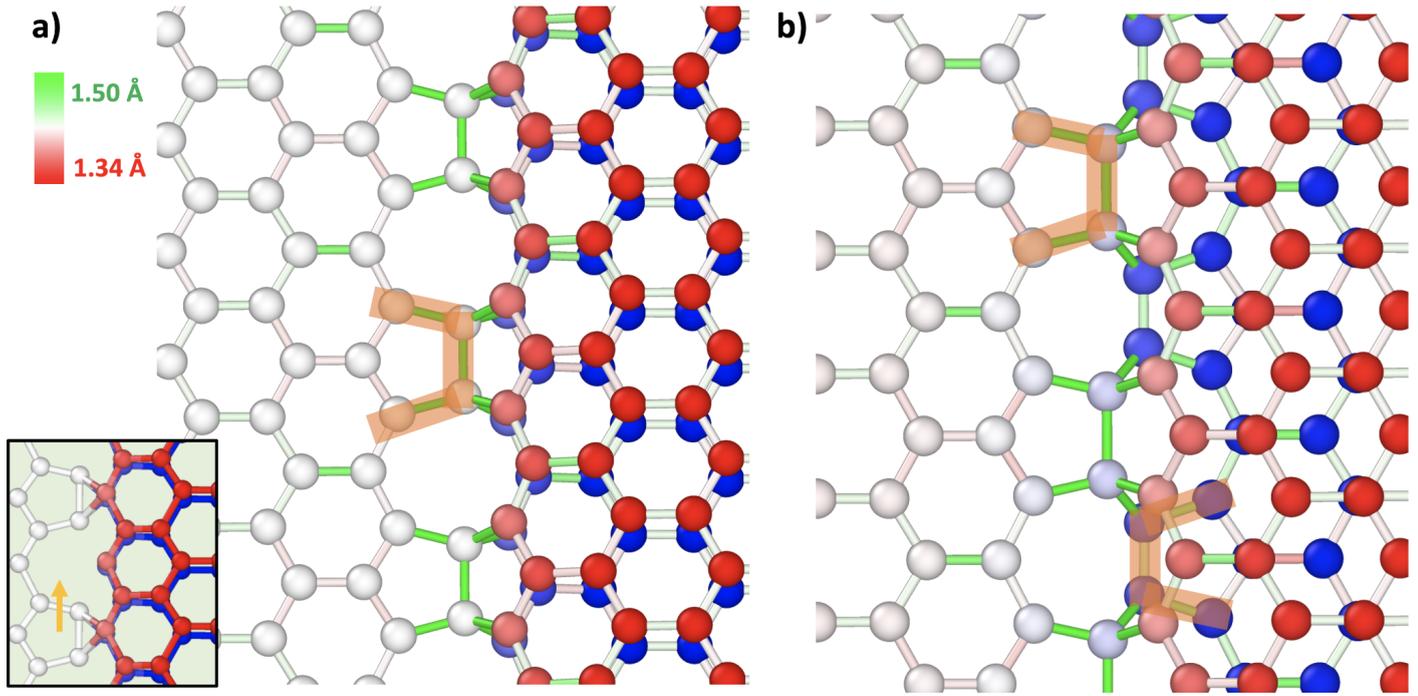

**Fig. 7: a)** RK-ZZ-ZZ core, featuring one pentagonal bond reconstruction in the ML region, inset shows the initial core before shearing in the y-direction and **b)** RK-ZZ-RK core, which allows AB stacking. Reconstructed pentagons are indicated for clarity.

### 3.1.3 Interlayer-bonded Armchair Cores

For armchair ILB configurations, there are only two distinct structures to consider, i.e. with or without a shift in one bilayer, giving either AA- or AB-stacking respectively (**Table 2**). The lowest energy configuration is the AA-stacked AC-AC-AC prismatic core (**Fig. 8a**) i.e. an AC-armchair edge bonding into the layer below, which is substantially lower in energy than the armchair free-standing edge, and has the lowest overall line energy of all cores considered in this work, again in good agreement with previous theoretical and experimental results [12,19,21]. The AC-AC-AC core results in moderately more strain than the ZZ-ZZ-ZZ core, with C-C bonds along the core direction showing local double bond character. In contrast, the AB-stacked core is substantially less stable. While the zigzag direction can generate AB stacking through only one lateral translation, the armchair core requires both a lateral and a screw component. This results in material adjacent to the core shearing away from AB stacking in the equilibrium geometry, as seen in **Fig. 8b**. In summary, and in contrast to Ref [21] we find that armchair-oriented dislocation cores are most stable when AA-stacked with interlayer bonding. The AB core is higher in energy, with FSE and ILB structures that are almost equal in energy.

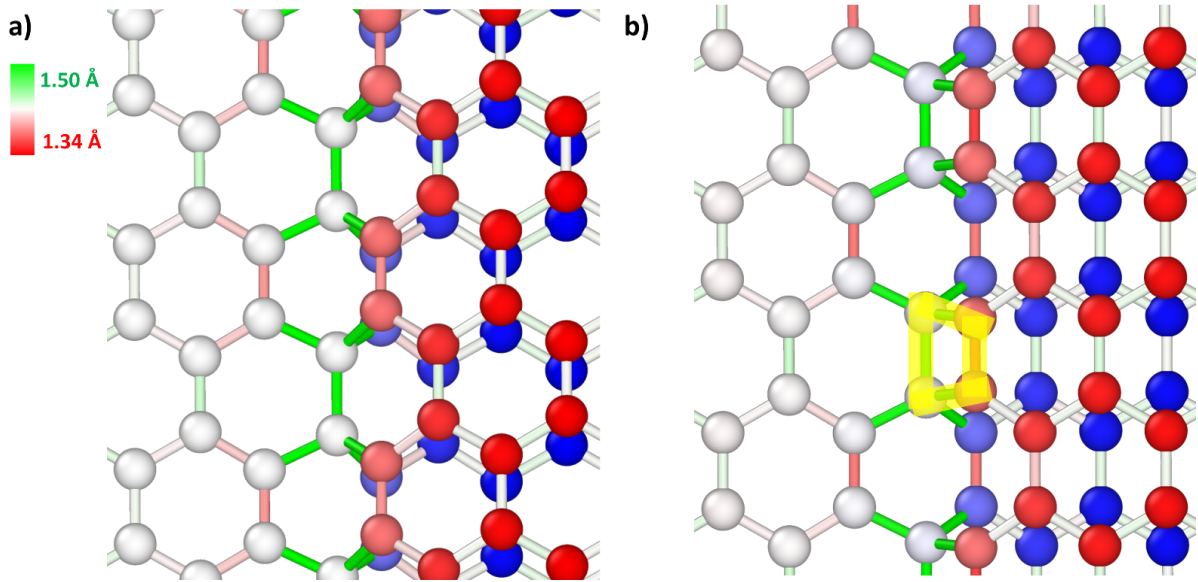

**Fig. 8**: **a)** AC-AC-AC dislocation core structure **b)** AC-AC-AC' dislocation core structure (square structure marked in yellow as a guide to the eye). C-C bonds are coloured depending on the degree of bond strain as shown in the key on the left, with white matching the pristine graphene bond length (1.42Å).

Finally, we note that our calculations suggest the possible importance of the local stacking environment to the type of reconstructions which are available. For example, while the ZZ-ZZ-ZZ-α ILB is almost equal in energy to the FSE-57 structure, this is not the case for AB-stacking, where all ILB configurations are much higher in energy. Detailed consideration of the stacking around a dislocation core will then be particularly important in the bulk of graphite. The lower formation energy of cores which give AA-stacking must be weighed against the additional energetic cost of faulted material adjacent to the core, which may restore AB-stacking (rhombohedral in bulk), at the energetic cost of a basal dislocation. As the latter energy is relatively low, at approximately 0.07 eV/Å [45], it is therefore likely that the AA-stacked cores are preferred in reality. Incorporating this energy changes the relative stability of the cores and gives us a stability order for the most stable cores to be AC-AC-AC < FSE AC < RK-ZZ-ZZ < FSE 57 < ZZ-ZZ-ZZ-α, all with relatively close energies (within 0.06 eV/Å).

### 3.2 Bulk Graphite

Having discussed the structure and energetics of isolated cores, we now consider different possible bulk reconstructions. When performing simulations of dislocations in solids, due consideration must be given to the unit cell and boundary conditions employed. There are many inequivalent dislocation arrays which can be used to construct the type of periodic supercells required for DFT simulation [46]. The two configurations of dislocation arrays used in this work are shown schematically in **Fig. 9**, corresponding to a quadrupolar array (**Fig. 9a**), where the Burgers vector alternates perpendicular to the line direction of the dislocation core, and a grain boundary configuration (**Fig. 9b**), where the dislocations separating regions of rotated graphite structure are all of the same sign (see also **Fig. 9e-g**). We have created quadrupole dislocation loops in bulk supercells through the insertion of extra material into an otherwise perfect graphite supercell (**Fig. 9a**) (equivalent to cleaving a perfect graphene sheet along the appropriate direction). The resultant orthorhombic cell contains a dislocation quadrupole, and through judicious use of appropriate boundary conditions, it is possible to construct an identical network using fewer atoms within a monoclinic cell as shown in **Fig. 9c,d**. [46, 47]

All cells have been constructed such that the net Burgers vector within them is zero (some typical examples are shown in **Fig. 9c,d**), and the lowest energy configuration of every cell is therefore perfectly-stacked, Bernal AB graphite. The creation of a prismatic defect then requires some formation energy to nucleate. Due to cell size limitations and boundary conditions for the grain boundary cells, we have inserted prismatic cores into graphite cells which are initially AA-stacked. During optimization the graphene layers easily shear with respect to one another due to strain around the core, and the final configuration is close to perfect AB stacking in the non-defective regions away from the core. For this reason, we have calculated all formation energies with respect to the energy per atom in the AB stacking configuration. This also provides an absolute upper limit on the formation energy in bulk, as AB stacking is the global minimum of perfect graphite.

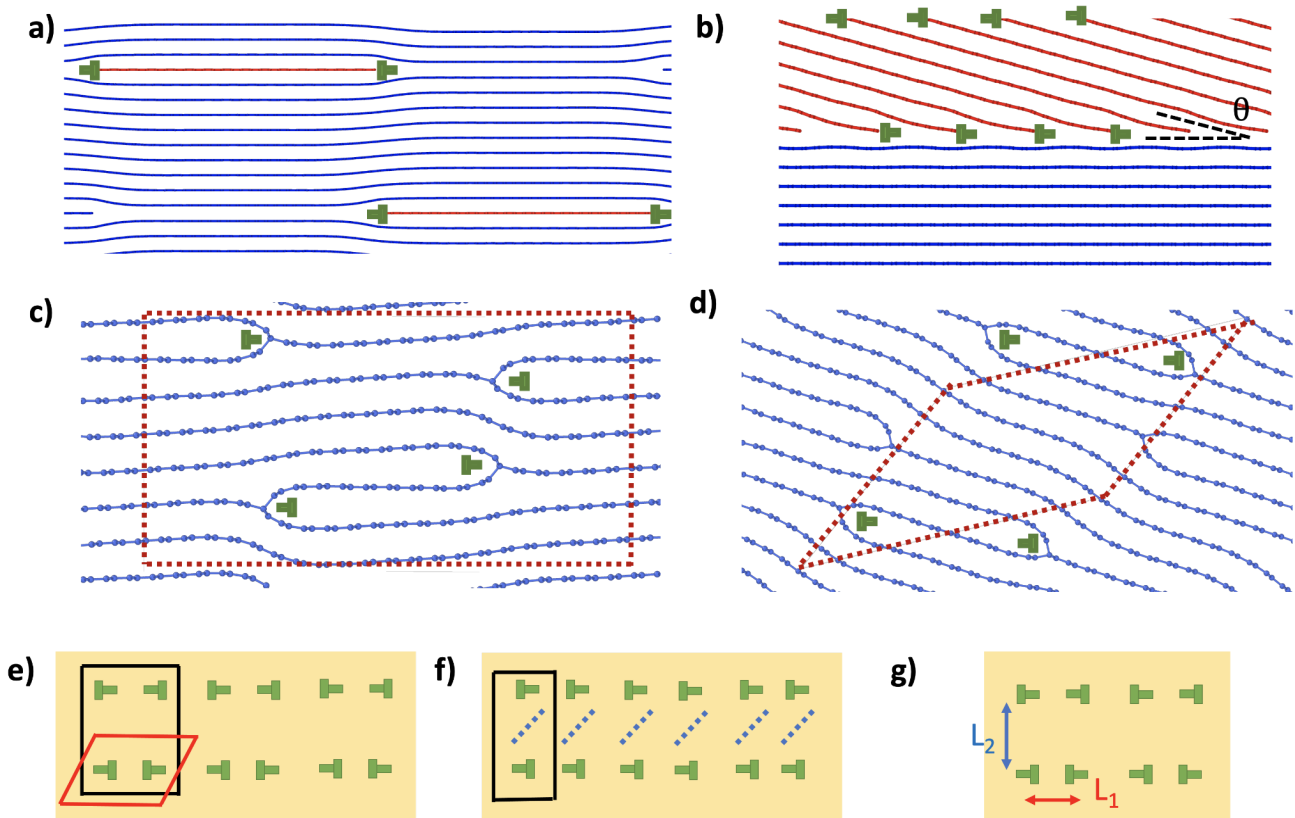

**Fig. 9:** Schematics of different simulation cells. **a)** Prismatic loops in bulk graphite, each nanoribbons inserted into the bulk giving a pair of prismatic dislocations. **b)** A prismatic dislocation array at a graphite c-axis tilt grain boundary. **c)** Relaxed supercells containing a quadrupole configuration of zigzag dislocation cores in an orthorhombic supercell and

**d)** a monoclinic supercell. **e)** Schematic depiction of the quadrupole configuration in orthorhombic (black) and monoclinic (red) simulation cells. **f)** Schematic of the grain boundary configuration. **g)** Structural parameters used to label dislocation arrays.

### 3.2.1 Reconstructions and stacking in bulk graphite

In order to evaluate different structures, we initially compare the line energy of different core reconstructions in a monoclinic cell, for both zigzag and armchair directions with approximate dimensions of 23 Å × 25 Å for both . Additional calculations using different boundary conditions have been performed and are briefly reviewed in the **Supplementary Material Section 2**, the cell structure will not have a large impact on results for the cell sizes considered here. Note when comparing bulk and isolated structures, the "ML" and "Bottom" edges from the isolated case can be considered as belonging to the same single graphene layer, while the "Top" sheet contains the bonding-in prismatic sheet edge. The layers within the bulk are explicitly labelled to clarify this point in **Fig. 10a**, and this notation is used in labelling different core configurations. A number of structures are considered based on the isolated core results, including the FSE-57, ZZ-ZZ-ZZ in α and β configurations and RK edges with and without pentagonal reconstruction, and both AC-AC-AC and AC-AC-AC' cores, in all cases considering both ILB and FSE configurations.

**Table 5** shows the line energy of the different prismatic cores calculated in this way. We note a net increase in the formation energy of all cores compared to the corresponding isolated structure, due to the additional elastic energy created by a lattice of prismatic dipoles in bulk graphite. The armchair cores remain the lowest-energy of all terminations examined in this work. Direct comparison between isolated and bulk cells suggests an approximate elastic energy of 0.384 eV/Å and 0.363 eV/Å for the zigzag and armchair cores, respectively. These are in quite close agreement with values calculated using elasticity theory for a superlattice of dipoles (0.45 eV/Å, 0.35 eV/Å for zigzag and armchair respectively) [21].

| *ZigZag Line energy (eV/Å)* | | | | | |
|---|---|---|---|---|---|
| ML | Top | Bottom | Type | $E_f$ | Δ |
| ZZ | 57 | ZZ | FSE | 1.323 | 0.000 |
| ZZ | ZZ | ZZ | ILB | 1.409 | 0.086 |
| ZZ | ZZ | ZZ | ILB-β | 1.593 | 0.270 |
| ZZ | ZZ | ZZ | FSE | 1.632 | 0.309 |
| ZZ | RK | ZZ | ILB | 1.672 | 0.349 |
| ZZ | ZZ | ZZ | FSE-β | 1.675 | 0.352 |
| ZZ | RK | ZZ | FSE | 1.702 | 0.379 |
| RK | ZZ | ZZ | ILB | 1.897 | 0.574 |
| ZZ | RK | ZZ | ILB* | 2.172 | 0.849 |
| ZZ | RK | ZZ | FSE* | 2.212 | 0.889 |
| *Armchair Line energy (eV/Å)* | | | | | |
| ML | Top | Bottom | Type | $E_f$ | Δ |
| AC | AC | AC | ILB | 1.263 | 0.000 |
| AC | AC | AC' | ILB | 1.429 | 0.165 |
| AC | AC | AC | FSE | 1.439 | 0.176 |
| AC | AC | AC' | FSE | 1.382 | 0.119 |

\* These are the ZZ-RK-ZZ cores without a pentagonal reconstruction.(see **Fig. 11b**)

**Table 5:** Formation line energies, $E_F$, (eV/Å), for both interlayer bonded (ILB) and free-standing edge (FSE) armchair and zigzag cores in a bulk monoclinic cell. **Δ** gives the difference in line energy between each configuration and the lowest-energy. Low-energy ILB ZZ cores are highlighted in green and are shown in **Fig. 10,** AC are highlighted in yellow and shown in **Fig. 12**)

Bond structure around the core is very similar between the isolated and bulk cores, shown in **Fig. 10** for the three lowest energy ILB structures. In both crystallographic directions, we find relatively low energy ILB configurations. For the zigzag direction, we again find that the non-bonded FSE Dienes reconstructed edge is preferred over ZZ-ZZ-ZZ ILB with a moderate increase in the energy difference from 0.03eV/Å to 0.08 eV/Å. The α-ILB reconstruction remains energetically preferred over β-ILB, although the difference is significantly reduced to 0.18 eV/Å, which we partly attribute to reduction in the bond angle compared to the isolated case due to the surrounding layers. In the armchair core, we find that the ILB structure is the lowest energy overall which is in contrast to an earlier report for the armchair edge, where it was found that the free-standing armchair edge was more stable. [21] This could be due to differences in cell construction, or due to computational approximations, such as the use of a Gaussian rather than plane-wave basis.

.

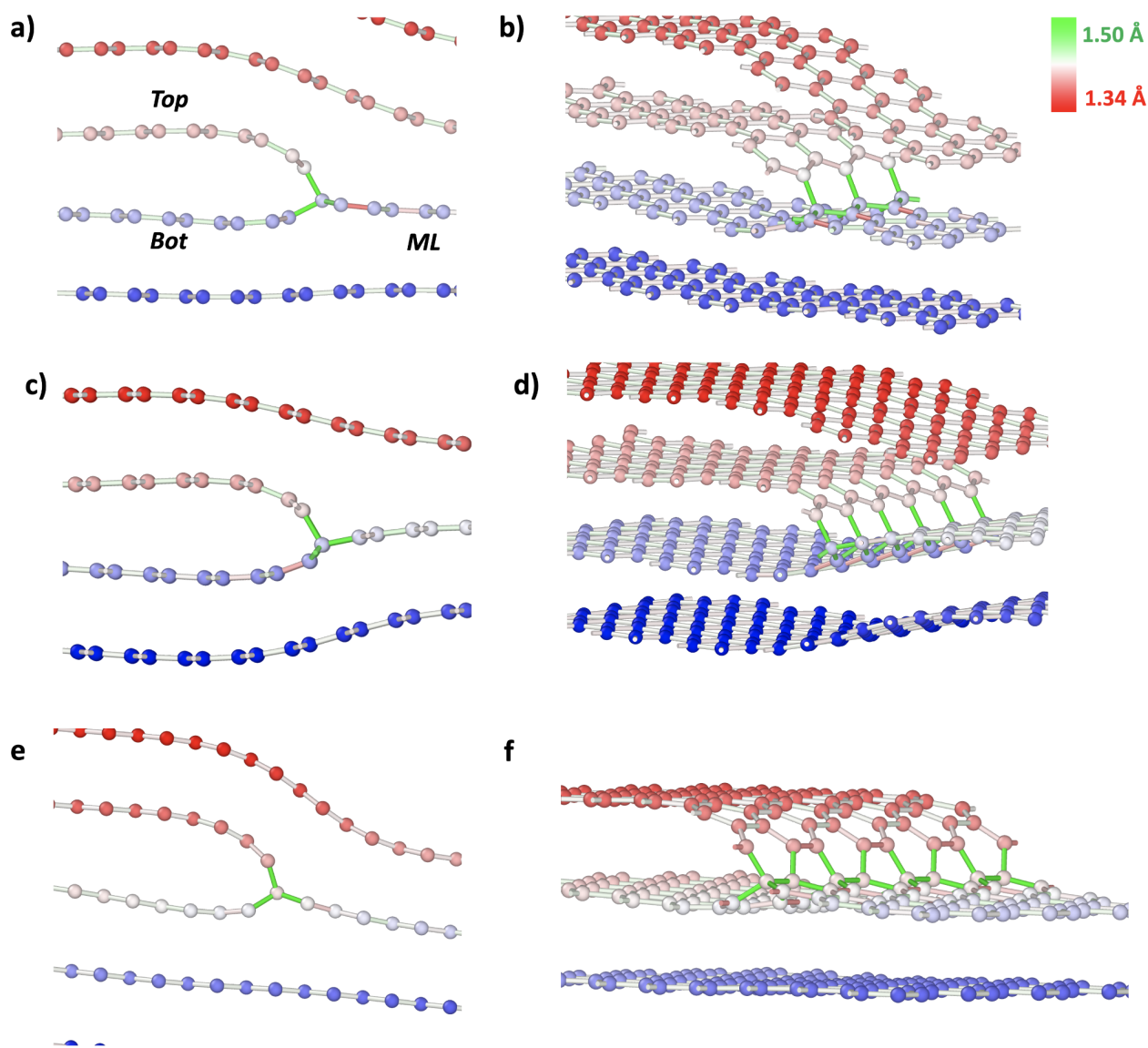

**Fig. 10:** Side and end-views of the **a,b)** zigzag (α-ZZ), **c,d)** β-ZZ and **e,f)** armchair (AC-AC-AC) ILB dislocation cores in bulk graphite.

ZZ-RK-ZZ cores, created through removal of a line of atoms in the "top" or bonding-in layer followed by different reconstructions (**Fig. 11a-c**), all have significantly higher energies compared to the ILB ZZ (**Fig. 10 a,b**) and Dienes reconstructed edges (**Fig 11d**). We have also considered the RK-ZZ-ZZ dislocation core (that is, an RK and ZZ edge *within the same graphite layer*). While this structure was more stable than the ZZ-ZZ-ZZ core for an isolated core, it would require the removal or addition of a row of carbon atoms within a layer in the bulk, in addition to a screw component which generates extensive additional strain which is not evident for the isolated cells due to lower dimensionality. In the bulk this gives a relatively distorted core structure with a high line energy of 1.89 eV/A, which is higher in energy than all examined reconstructions apart from the unreconstructed ZZ-RK-ZZ core (**Fig. 11b**).

Thus for the zigzag orientation we have an interesting situation, whereby the most stable zigzag oriented prismatic core is the ZZ-57-ZZ, non-bonded, with the dislocation line edge reconstructed through bond rotations to give a pentagon-heptagon edge sequence with triple-bond character at edge bonds (**Fig. 11d**). However close in energy is the ZZ-ZZ-ZZ interlayer bonded core, where an unreconstructed zigzag edge bonds directly into the sheet below. These two structures may potentially result in an interesting mechanical response that depends on annealing and restructuring at the core, and is discussed further below.

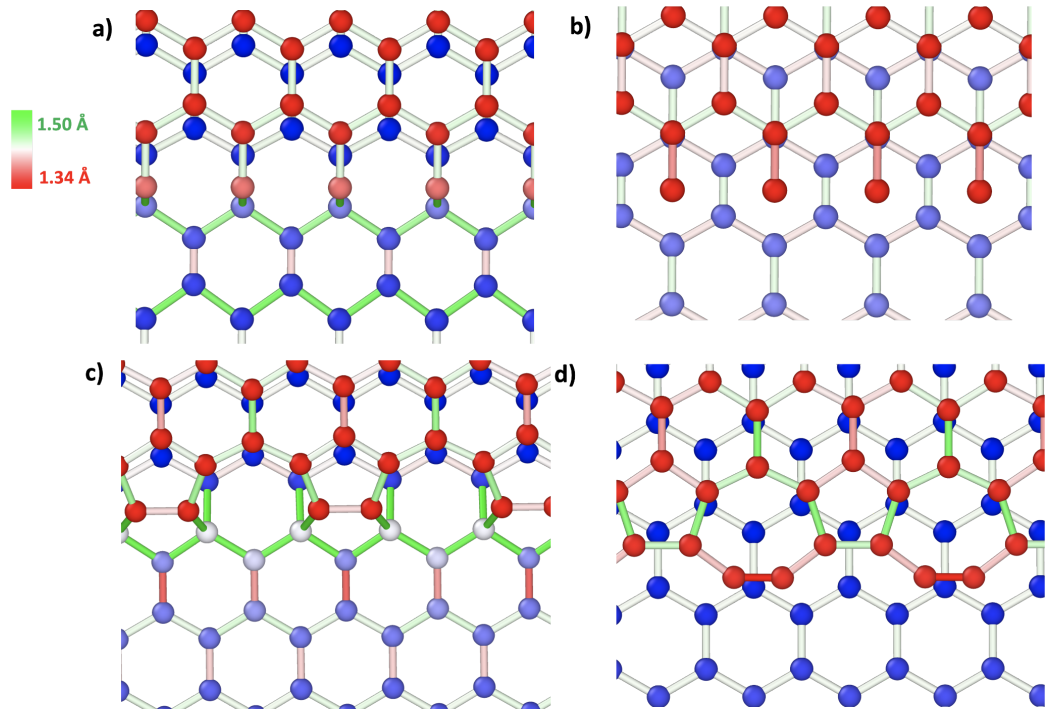

**Fig. 11:** Different bulk zigzag prismatic configurations, calculated in a monoclinic cell **a)** Unreconstructed ZZ-RK-ZZ in an ILB and **b)** FSE configuration **c)** ILB ZZ-RK-ZZ core with pentagonal reconstruction. **d)** The lowest-energy ZZ-57-ZZ core. Bonds that are stretched (compressed) are coloured green (red).

Turning now to the armchair case, we consider both the purely hexagonal AC-AC-AC and the less stable 4-8 AC-AC-AC' reconstruction. Similarly to the difference between the α and β cores, we can create both structures in the bulk by shifting the adjacent flat sheet by a bond length, as indicated in **Fig. 12** by arrows, showing how the breaking of bonds and shifting of layers creates these different cores. The FSE configurations have an almost identical line energy and structure, and are slightly less stable than the ILB structures. The shifted AC' core is also higher in energy than the AC core, due to the bonding strain from 4-8 defects, and is much closer in energy to its corresponding FSE structure than the shifted AC core.

Thus in the armchair case we find that bonding into the layer below is the most stable configuration, with relatively little energy difference when the layer debonds and becomes free-standing. In general the armchair oriented cores are slightly more stable than zig-zag oriented cores.

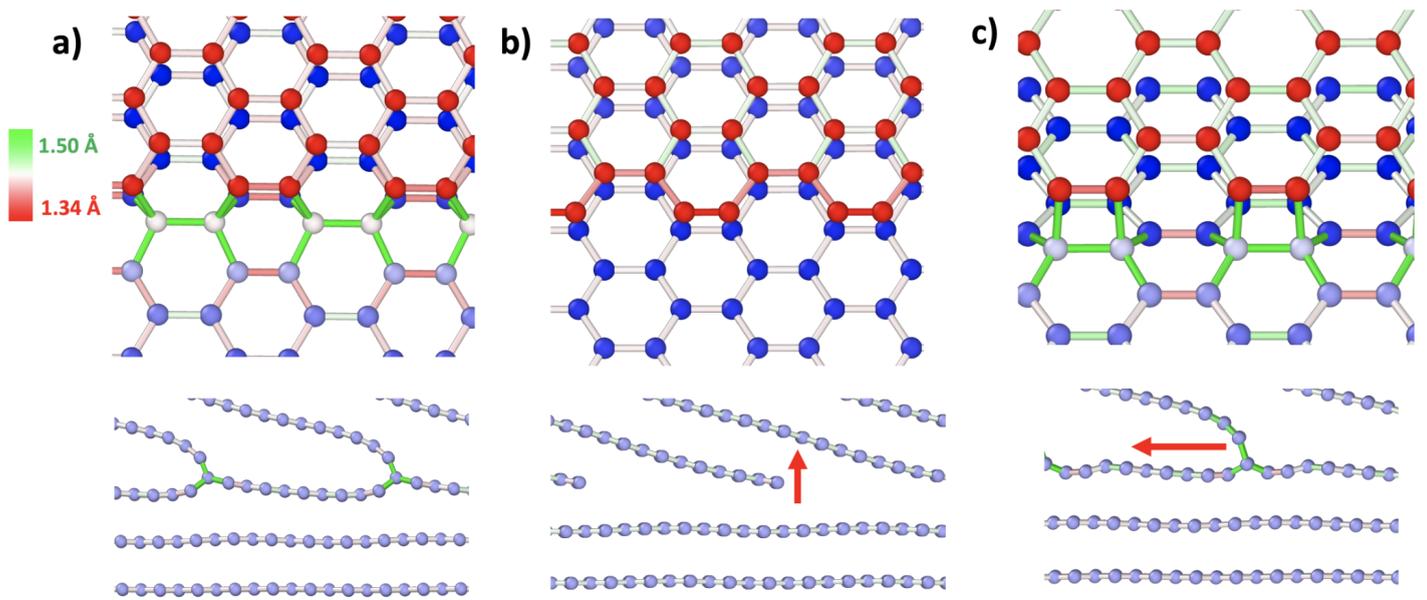

*Fig. 12: Armchair prismatic core configurations **a)** Initial hexagonally-bonded configuration (AC-AC-AC ILB) **b)** AC-AC-AC FSE **c)** AC-AC-AC', which features the 4-8 square structure.*

### 3.2.2 Grain boundary cells

We now consider asymmetric c-axis tilt grain boundaries. These consist of two graphite grains, one of which lies flat with θ = 0, while the other has been tilted in the c-axis direction such that θ > 0. This gives an array of graphene edges (prismatic dislocations) where the tilted grain meets the flat grain in the boundary region (**see Fig. 9b**). These cells have been created using appropriate boundary conditions where the atoms bordering one cell edge (i.e. the right-hand side) in one layer are continuously connected to the left-hand side atoms of the layer above (**Fig. 13b**).

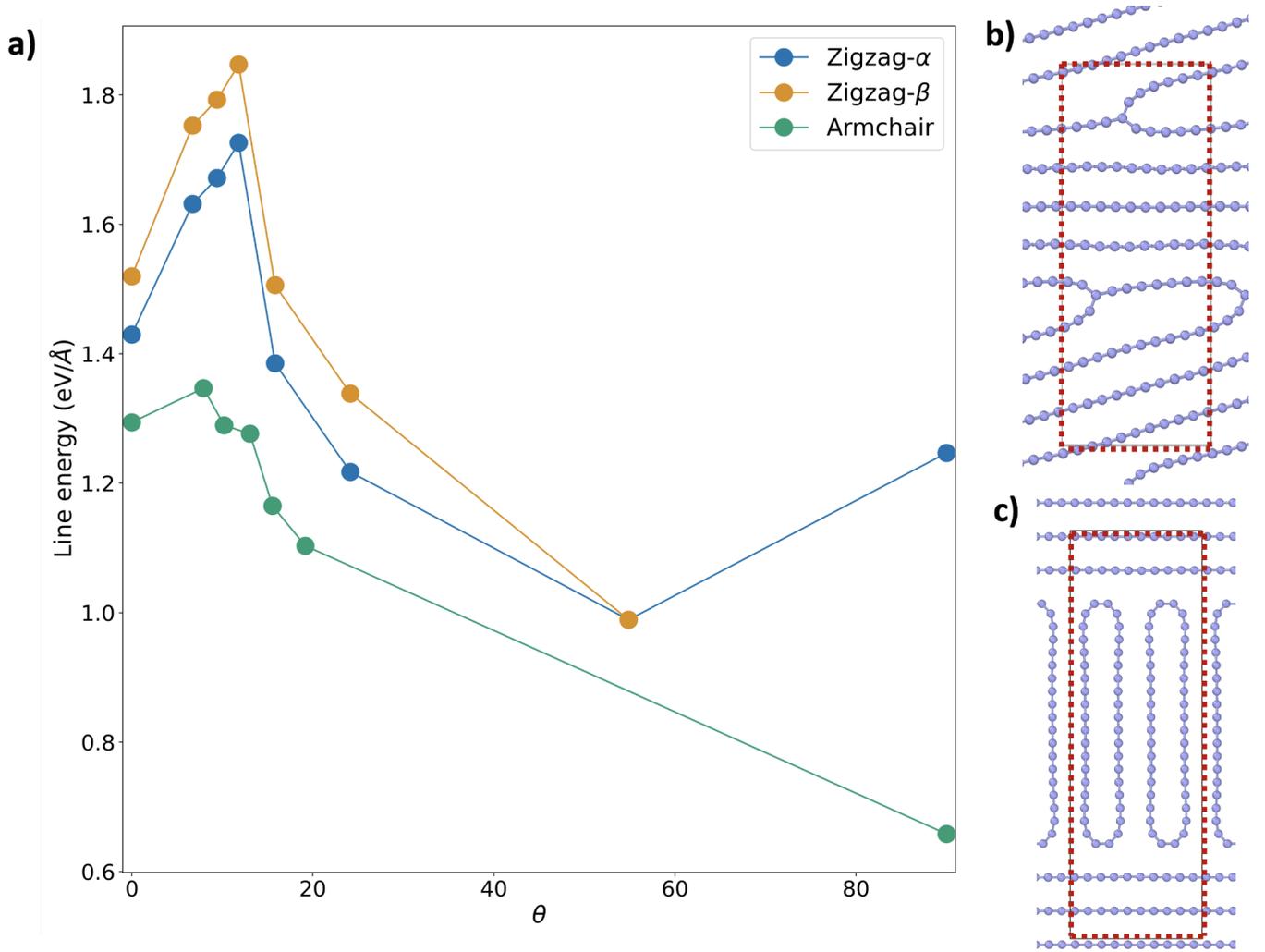

**Fig. 13:** Fully-relaxed prismatic dislocation quadrupole cells **a)** Formation line energy vs angle for the armchair and zigzag cores, 0° corresponds to the quadrupole/isolated value. **b)** 19° armchair grain boundary **c)** $sp^2$ capped 90° armchair grain boundary.

As the inter-grain tilt varies, the geometric incompatibility between tilted regions results in different densities of graphene edges along the grain boundary. This results in the tilted graphene sheets being "cut" with a specific crystallographic character at the grain boundary. Using simple geometric considerations, the separation (d) between carbon atoms along a prismatic edge at an asymmetric c-axis tilt grain boundary is

$$d = \frac{c}{\sin\theta}.$$

Graphite c-axis grain boundaries change the angle of the interlayer $sp^3$ bonds, and naturally create a mismatch in the lattice parameters between the flat and tilted grains. These two effects modify the formation energy of a prismatic array at the boundary. The DFT cells used in this work are necessarily small, and we stress that this could influence conditions of strain accommodation at the grain boundary. Nonetheless, we are able to observe some broad trends in our calculations.

**Fig. 13a** shows the formation energy of a prismatic dislocation as a function of angle, for the armchair and zigzag core in both α and β-orientations. At large angles (close to 90°), the terminating edge bonds are close to each other and their separation is on the order of the interlayer separation. This allows the prismatic edges to form perfect graphitic bonds between adjacent edges, resulting in the $sp^2$-capped configurations shown in **Fig. 13c**, which our calculations confirm are lower in energy than the corresponding $sp^3$ interlayer bonding. This type of looped closed-edge is readily observed at graphite crystallite boundaries in TEM experiments. [48-50]

At smaller angles the bond separation increases and the density of bonds along the boundary decreases, such that there are no adjacent uncompensated bonds for the edge atoms to form a perfect $sp^2$ termination. These edge structures then reduce their energy through interlayer bonding, so that the "edge capping" now effectively occurs via bonding into the adjacent layers, resulting in an array of prismatic dislocations. Interestingly, there is a marked change in the energy dependence for angles below ~15°, where we find a larger curvature in the flat layers immediately adjacent to the dislocation core (for additional images (see **Supplementary Material Fig. S2**). At these low tilt angles, the quadrupole core energies (i.e. dispersed dislocations) are lower than along a grain boundary. Typically, migration of dislocations towards a grain boundary can significantly lower defect formation energy, and our calculations indicate that this process may be inhibited for low-angle tilt grain boundaries which could lead to incoherent interfaces between graphite grains. Above 15°, grain boundary configurations are lower in energy than the quadrupoles and will tend to align, similar to dislocation cores in silicon [47]

Notably, our calculations at an intermediate angle of $\theta \approx 60°$ reveal a very low core energy, and no clear preference for either the α or β ILB cores, with tilted graphene edge atoms bonding into different atoms in the flat grain (see **Supplementary Material Fig. S3**). The $sp^3$ bond angle remains close to 90° at these angles, and the grain tilt arises from net curvature of the adjacent graphene sheet. We can understand this by noting that, in general, there is an energetic competition at play in the angled grain boundary interfaces. On the one hand, the $sp^3$-bonded carbon atoms in the tilted grain will prefer a 90° contact angle, which drives the tilted planes towards attaching orthogonal to the flat sheet. However, a 90° match is not favoured when matching the interlayer spacing in the tilted grain to the in-plane periodicity of the flat grain. In practise, for large-angle grain boundaries such as the 60°, this results in a configuration where the layer spacing matches, with local curvature in the layers at the boundary towards a 90° contact angle (see **Fig. S2**).

We also notice that these interlayer configurations can localise $sp^2$ bonds. At these angles, arrays of dislocation cores effectively create lines of internal surfaces, which show similar behaviour to an external reconstructed non-functionalised diamond surface, so-called "Pandey π-bonded chains". [51,52] This is exemplified by the two parallel zigzag chains, where the "external" atoms are $sp^3$ with only single bonds. This results in pairs of atoms with a spare $p_z$ orbital, which are completely isolated from all other $sp^2$ atoms in the lattice, so that they form a strong π-bond together, as shown in **Supplementary Material Fig. S3,** where they are joined by green bonds due to substantial bond compression.

## 3.3 Dislocation migration and thermal stability

Thus far the discussion has focussed on the thermodynamic stability of prismatic dislocation core structures. Processes of bond formation, breaking and reconstruction within a core region are among the most fundamental properties of the prismatic edge dislocation core, and the kinetics of dislocation motion is a key factor in determining structure and properties. Notably, while certain structures may be thermodynamically favoured, kinetically they may not be easily accessible. Additionally, bond-breaking processes are paramount both in determining the relative thermal stability of ILB configurations and in determining dislocation mobility. Prismatic dislocation motion can occur along two inequivalent directions: between different basal planes, and therefore *along* the Burgers vector direction (i.e. in the c-axis direction) via dislocation *glide*, and along a basal plane or *perpendicular* to its Burgers vector, via dislocation *climb*. Dislocation glide must proceed by bonding into a different layer, through the successive debonding and rebonding of the carbon atoms along the dislocation line in a process of double-kink formation followed by kink propagation, which necessitates the breaking of interlayer bonds within the core region.

To address these questions, CI-NEB calculations have been performed between the ILB (bonded) and FSE (broken) configurations of the bulk zigzag (α-ZZ) and armchair (AC) core configurations. Initially, we work with the same thin slab unit cells used in our energetic calculations, and so we are obliged to break all bonds along the dislocation core simultaneously. This means that our results do not incorporate the formation and propagation of kinks (in glide) and jogs (in climb) respectively, which are beyond the scope of the current study. These processes are likely to be important in realistic physical scenarios, and so these results should be taken purely as an indication of the most important physical processes mediating dislocation motion. We also consider the energetics of the bond breaking process in isolated cores, which have been created by extending bulk cores along the dislocation line.

### 3.3.1 Bond-breaking and glide

**Fig. 14a** shows the energy barrier vs transition coordinate of the debonding process for both the zigzag and armchair configurations and, in the inset, the initial and final structures of the NEB calculation. For the zigzag core, the energy barrier to break an interlayer bond is 0.56 eV in our ribbon cells (0.56 eV/bond or 0.23 eV/Å), whereas the reverse rebonding process costs only 0.218 eV (0.109 eV/bond or 0.054 eV/Å). The armchair core demonstrates a similar, though less pronounced, asymmetry, with forward and reverse barriers of 1.478 eV (0.739 eV/bond or 0.346 eV/Å) and 0.676 (0.338 eV/bond or 0.158 eV/Å), respectively.

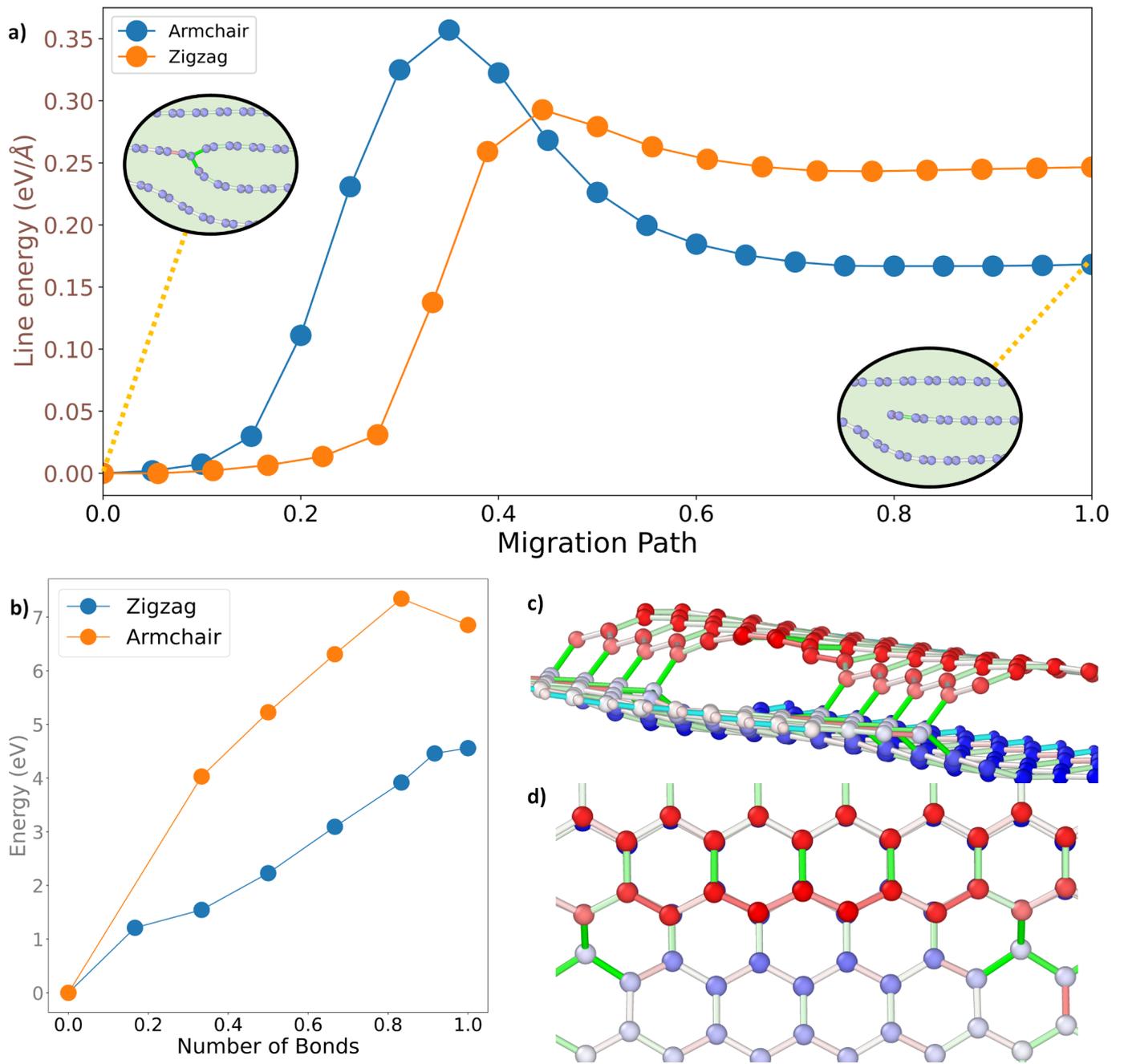

**Fig. 14: a)** Energy vs transition coordinate for the spontaneous breaking process of the zigzag (α-ZZ) and armchair (AC) prismatic cores **b)** Total energy fraction of broken bonds for an isolated zigzag (α-ZZ) prismatic core of length 1.9nm (12 interlayer bonds). This process remains endothermic and extensive in the number of interlayer bonds broken along the dislocation line. For n = 2 broken bonds along the armchair core and n=1 along the zigzag, we find that the core spontaneously reconstructs, and so this point has been omitted. **c)** Side and **d)** Plan view of broken bonds along the α-ZZ core.

On the basis of these first-principles energy barriers, we can estimate the probability that a single individual bond will fully break at a given temperature and infer the approximate temperature dependence of dislocation mobility. A minimal requirement for the formation of dislocation kinks will be the breaking of some fraction of the interlayer bonds along a dislocation line. We assume that the thermal activation of an individual interlayer bond is completely uncorrelated from that of adjacent bonds, and ignore conditions of in-plane and interlayer strain, and kink nucleation. We anticipate that strain effects will tend to increase the energy barrier while the nucleation and propagation of kinks can lower the required temperature since it is necessary for fewer bonds to break. Hence, the following calculation is a rough order of magnitude estimate only of the dislocation mobility.

The transition rates between the broken and unbroken configurations are evaluated using the Arrhenius equation, which is appropriate under the assumption that the timescales of bond-breaking and thermal processes are widely separated. The transition rate between any two configurations is given by

$$r_{i \to j} = A_0 e^{-\frac{\Delta E_{i \to j}}{k_B T}}, \quad (1)$$

where $\Delta E_{i \to j}$ is the corresponding energy barrier between configurations **i** and **j**, $k_B$ is Boltzmann's constant, T is the temperature and $A_0$ is the attempt frequency, which is taken to have the value $A_0 = 10^{13}$ Hz. Given a set of transition rates and under that assumption of thermal equilibrium, the probability that a single interlayer bond is in any particular configuration at a given temperature is

$$p_i = \frac{\Sigma_j r_{j \to i}}{\Sigma_{i,j} r_{j \to i}}, \quad (2)$$

which we then evaluate using the respective DFT barriers heights. The resulting probability distribution (see **Supplementary Information Fig S5**) shows that at low temperatures (T<1500K) the probability that an individual bond is broken is essentially zero, implying a very low density of broken bonds along a dislocation line. Cores are highly thermodynamically stable and immobile up to relatively high temperatures (>1500K), above which there is an appreciable population of broken bonds at thermodynamic equilibrium. Although approximate, this order of magnitude estimate compares reasonably well with experimental imaging of prismatic cores in carbon nanotubes, where spontaneous prismatic glide is not seen at temperatures below 2000K. [16]

We note that this discussion does not consider the ZZ-57-ZZ FSE core, which in principle is the most stable ZZ prismatic core. This cannot glide in bulk without first reconstructing, since otherwise the glide process would involve incorporation of a 57-reconstructed edge into a fully hexagonal layer, which is highly unstable. This edge may come into play, for example through constrained annealing, but in the temperature range 1500-2000K we expect the edge will be able to reconstruct and hence seems unlikely in general to impede prismatic dislocation glide.

In the above discussion we assume that all bonds along the dislocation core break and reform during motion simultaneously. In practice however dislocation migration is likely, as in other materials, to commence with double kink formation, followed by kink migration. To model this process requires symmetry breaking along the dislocation core. In general this requires the use of extremely large supercells which are beyond the scope of this current study. However, to provide some insight into this process, we have performed a set of structural relaxations of the isolated α-ZZ core for different numbers of broken interlayer bonds (**Fig. 14b**). These calculations have been performed at the Γ-point, in an isolated cell which is periodic only in the direction of the dislocation line. Carbon atoms far from the core region have been fixed in single and bilayer configurations. and dangling carbon bonds have been passivated using hydrogen atoms. For additional details of the supercell used, see **Supplementary Information Section 6 & Fig. S6**)

We find that the bond-breaking process remains endothermic as a function of the number of broken bonds, with an average energy of approximately 0.4 eV/bond, which is larger than the nanoribbon cells due to conditions of strain and deformation along the dislocation line. The large and monotonic energy increase as a function of broken bonds is highly suggestive that processes such as kink and jog formation are required for dislocation motion. Local conditions of applied stress, the presence of environmental contaminants, edge reconstructions, and other processes are also likely to play an important role in realistic scenarios.

### 3.3.2 Annealing (Dislocation annihilation)

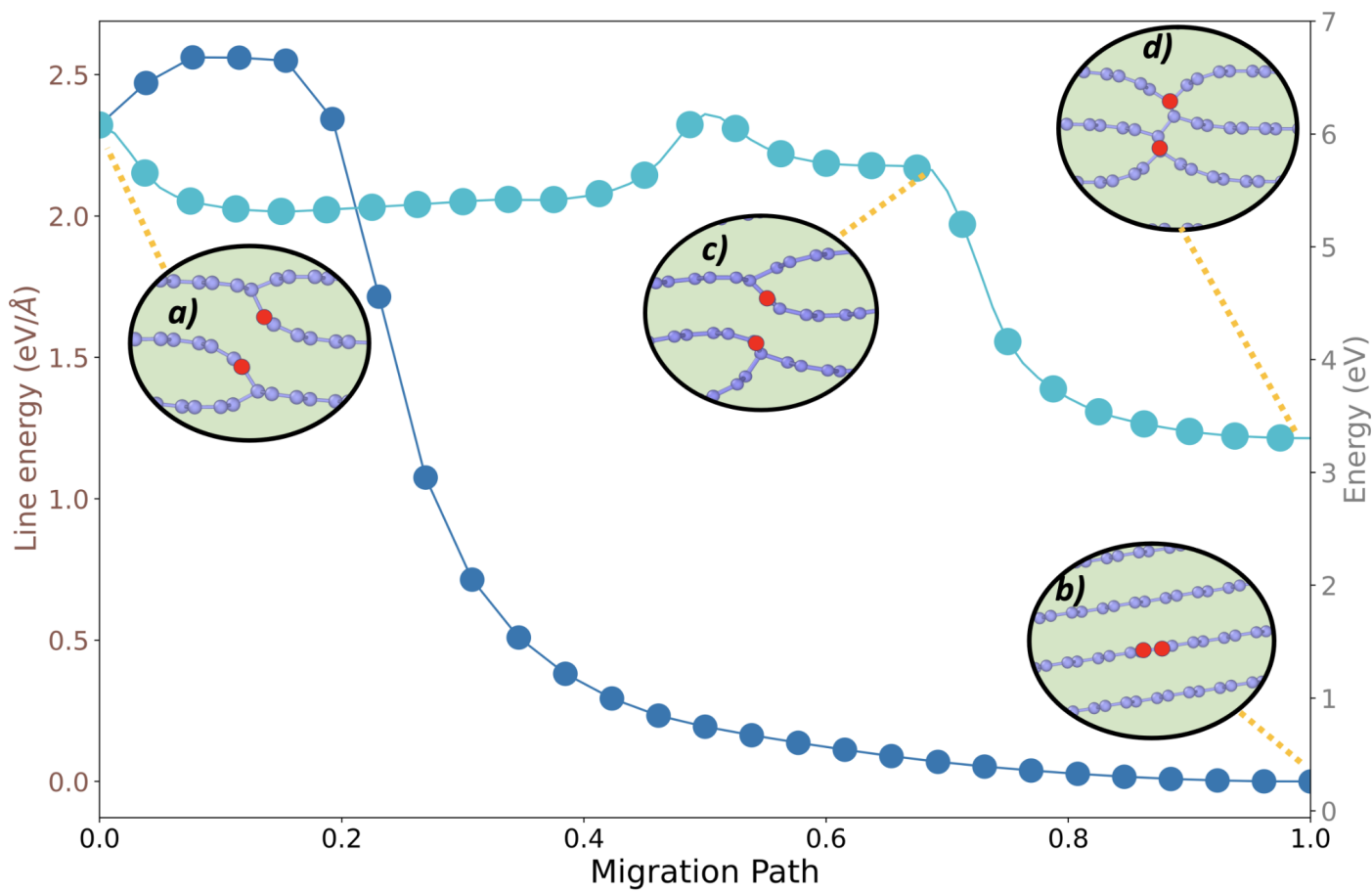

**Fig. 15:** Energy barriers to fully anneal (dark blue) and to form a metastable bound state (light blue) vs transition coordinate, from the same initial state. **a)** In order to fully anneal and restore perfect graphite structure, the two adjacent cores must adopt the same initial (α-ZZ) configuration, so that they have a net Burgers vector of zero, and the two bonded-in carbon atoms (highlighted, red) in each core lie on different sublattices. **b)** The adjacent pair of dislocations can then annihilate, resulting in three perfect graphene

sheets. **c)** The adjacent cores can also reorient i.e. they are unstable to a metastable reconstruction of one core towards a β-ZZ core. **d)** This structure can also substantially lower its energy through the formation of a single inter-core $sp^3$ bond.

Adjacent dislocations can strongly interact with one another, and dislocations of opposite signs should show strong, elastic-field driven attraction. At sufficiently high temperatures, this inter-core interaction necessitates the pile-up and eventual annihilation of oppositely-signed prismatic cores, a process which will release significant strain from the graphite lattice and anneal damage. We have examined this process using pairs of zigzag dislocation cores. In order to fully anneal and restore a perfect graphite lattice, it is necessary that both of the adjacent cores have the same (α-ZZ) core configuration. As annihilation must proceed through the breaking and rebonding of the edge structure in both adjacent cores, this guarantees that the carbon atoms along each zigzag edge inhabit opposite graphene sublattices, which is necessary to ensure perfect graphene structure in the reconstructed sheet. In principle this process can also occur for a pair of β-ZZ cores, although this is not considered here due to the higher energy of that structure.

**Fig. 15** shows the annihilation barrier and corresponding core structures of two oppositely-signed α-ZZ prismatic dislocations. The annihilation barrier of this dislocation pair (0.28 eV/Å) is only moderately higher than the barrier for normal zigzag glide, and the transition state for core annihilation is simply the FSE configuration of one edge, which is the same as the transition state for normal prismatic diffusion. Once the interlayer bonds of one core are broken, the corresponding bonds in the adjacent prismatic spontaneously break, leaving perfect $sp^2$ graphene sheets in the layers above and below, and joining the two half-sheets together and creating a new, perfect $sp^2$-bonded layer (**Fig. 15c**).

Notably, other low energy core structures are possible. For example, two adjacent α-ZZ cores can lower their energy by 0.3 eV/Å by converting one core to a β-ZZ structure through reorientation of the $sp^3$ bond (**Fig. 15c**). This structure can then form an $sp^3$-bridge between the adjacent cores, creating a fully $sp^3$-zone within a prismatic dislocation complex (**Fig. 15d**). This substantially lowers the energy compared to a pair of isolated dislocations, only requires a moderate energetic barrier to be circumvented and has a sufficiently high barrier for the reverse process that it is likely metastable. It is tempting to see such metastable structures in HRTEM images (see for example, the left circled region in Figure 19b, and the lower-right region in Figure 19c). It is likely that other metastable structures exist, and they may provide a partial explanation for the unannealable damage observed in high temperature irradiated graphite [53]. Such metastable structures lock in substantial amounts of energy into the lattice and may also be potential candidates for higher temperature energy release peaks in irradiated graphites.

### 3.3.3 Dislocation climb

Finally, we consider the complementary diffusion mechanism of dislocation climb. Prismatic climb is an integral part of the motion of the prismatic core, and it is not possible to fully consider the unconstrained motion of a prismatic core without it. Due to the Bernal stacking of the surrounding graphite lattice, a prismatic dislocation in the lowest-energy C-stacked configuration cannot glide indefinitely between layers, and reorientation of the prismatic edge so that the edge carbon atoms align with those on adjacent layer is necessary for completely free motion. This is shown schematically for the zigzag configuration in **Fig. 16**.

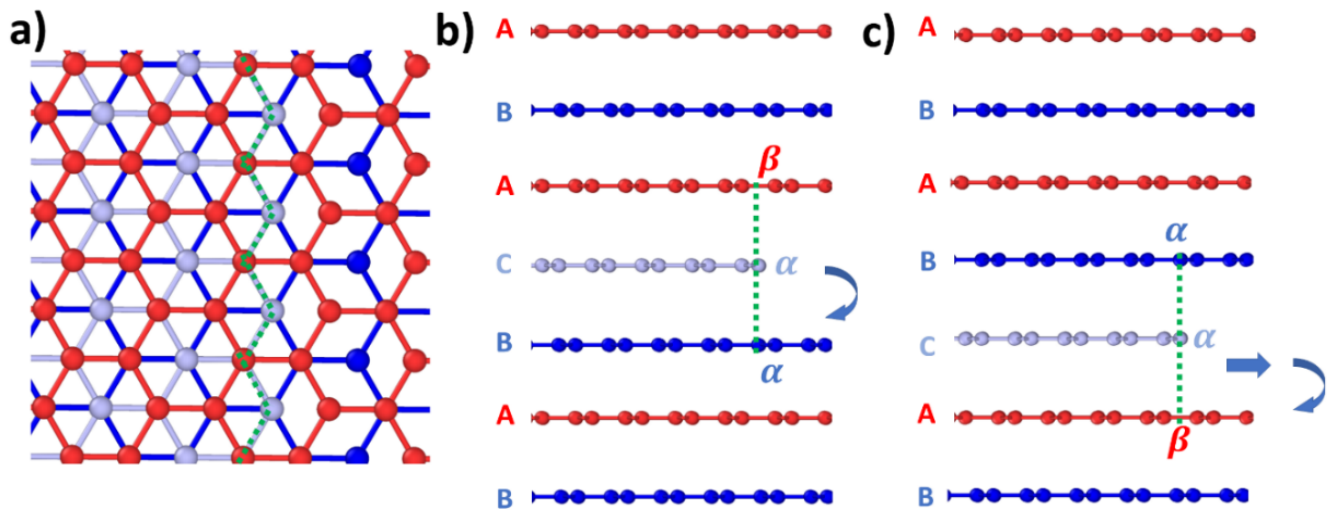

**Fig. 16:** Schematic depiction of prismatic glide and climb **a)** Top down view showing the initial structure of a prismatic core where the prismatic (C stacked) layer edge is on the same sublattice ($\alpha$ atom) as the lower, (B stacked layer) **b)** Side view of the same structure in a. The prismatic core can glide freely in this configuration between layers C and B. **c)** Further glide is inhibited because the prismatic edge atoms are above the hollow site of the A layer, this necessitates climb or interlayer sliding of the A layer.

While the prismatic core can easily bond into the adjacent B layer, due to the overlap of carbon atoms on the C and B layers, the A layer is distinct in that the edge atoms of the prismatic layer lie above a hollow site rather than a line of carbon atoms. Hence, in order to continue to perform glide in this direction, either the adjacent A-stacked sheet or the C-stacked prismatic sheet must *climb* - execute motion perpendicular to the c-axis direction - to facilitate further glide.

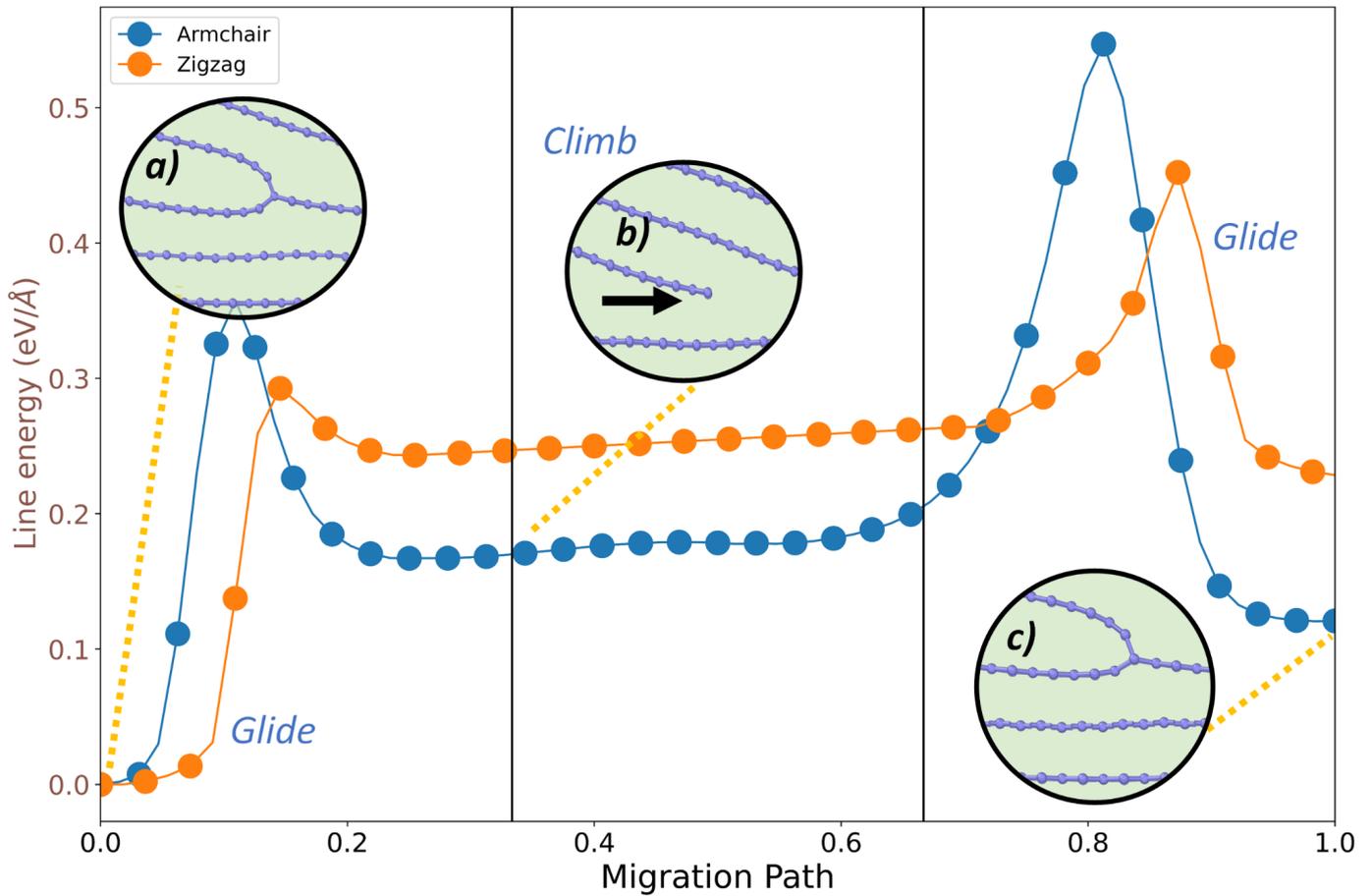

**Fig. 17:** *Prismatic climb barriers for the glide and climb processes, and (insets) armchair core configurations Note that the migration climb energy is dependent on the stacking fault width, which is small in these current cells, but is in general extensive and proportional to the area of stacking fault. **a)** Initial AC-AC-AC bonded core configuration. **b)** After bond-breaking, the core can climb by sliding adjacent graphene sheets. **c)** The re-bonding process which takes the core to the higher energy AC-AC-AC' square structure. (Only around half the calculated points are marked for visual clarity. Lines connect directly calculated points and are a guide for the eye.)*

The entire process is depicted for the armchair (zigzag) core in **Fig. 17**. A prismatic core which is initially bonded at the lowest-energy AC (α-ZZ) site, moves to an adjacent AC' (β-ZZ) site via a three-step barrier. The initial and final steps simply consist of the bond-breaking mechanism. The intermediate steps are $sp^2$-hybridised "dangling" edges, which climb simply by the passage of the flat bottom layer against the edge so as to alter the stacking. This process has a small barrier in our cell of 0.050 eV/Å. In reality, interlayer sliding is an extensive quantity, with an energetic penalty which is proportional to the area within a prismatic dislocation loop. In the bulk of graphite, this process is likely to proceed via the creation of a stacking fault and the emission of a basal dislocation, which as previously discussed are large and beyond the capability of conventional plane-wave DFT. However, our calculation does demonstrate the basic mechanism of prismatic climb, and is indicative that the climb – and therefore glide – should proceed relatively freely once interlayer bonds can be broken. Other processes, such interactions with vacancies and interstitials, are also likely to be relevant to prismatic climb.

### 3.4 Simulated TEM Images

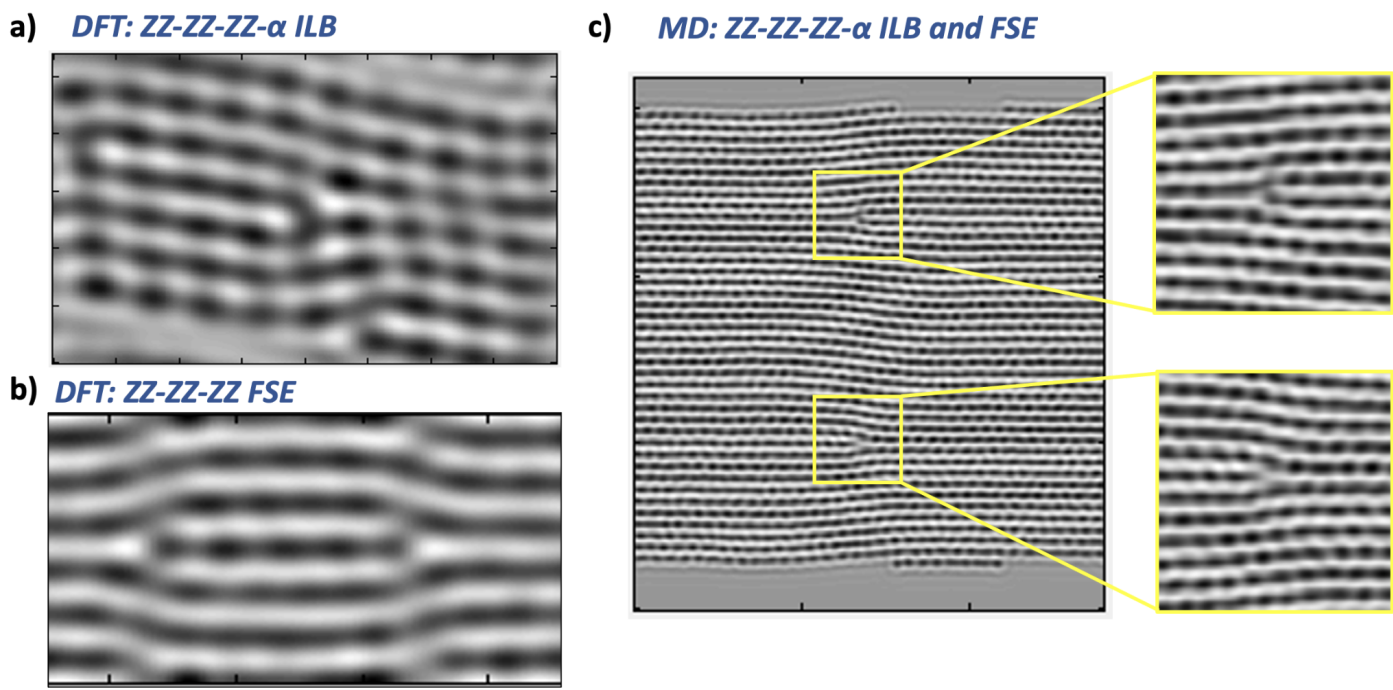

**Fig. 18:** Simulated TEM images of prismatic cores in graphite: (a) DFT optimized bulk monoclinic cell with ZZ-ZZ-ZZ-α ILB cores, (b) DFT-optimized bulk monoclinic cell with ZZ-ZZ-ZZ FSE cores, (c) Molecular dynamics bulk structure, optimized using LAMMPS [54] and the multilayer graphene hybrid neural network potential [55], having both (upper) ZZ-ZZ-ZZ-α ILB and (lower) ZZ-ZZ-ZZ FSE cores, as highlighted by yellow squares. Full defocus sequences are given in Supplementary Materials.

In order to determine the degree that different core structures will be visible experimentally, **Fig. 18** shows the results of simulated TEM images for ZZ-ZZ-ZZ ILB and FSE cores, from both smaller monoclinic DFT cells, and a larger MD simulation using LAMMPS [54] with the multilayer graphene hybrid neural network potential [55]. While the DFT structures can be clearly distinguished, the ILB is less evident in the MD simulated structure (top yellow square, **Fig. 18(c)**). The contrast for bonded reconstructed edges is also dependent on the degree of defocus (see Supplementary Materials), and this suggests that in non-$C_s$ corrected microscopes it may be difficult to distinguish FSE and ILB cores.

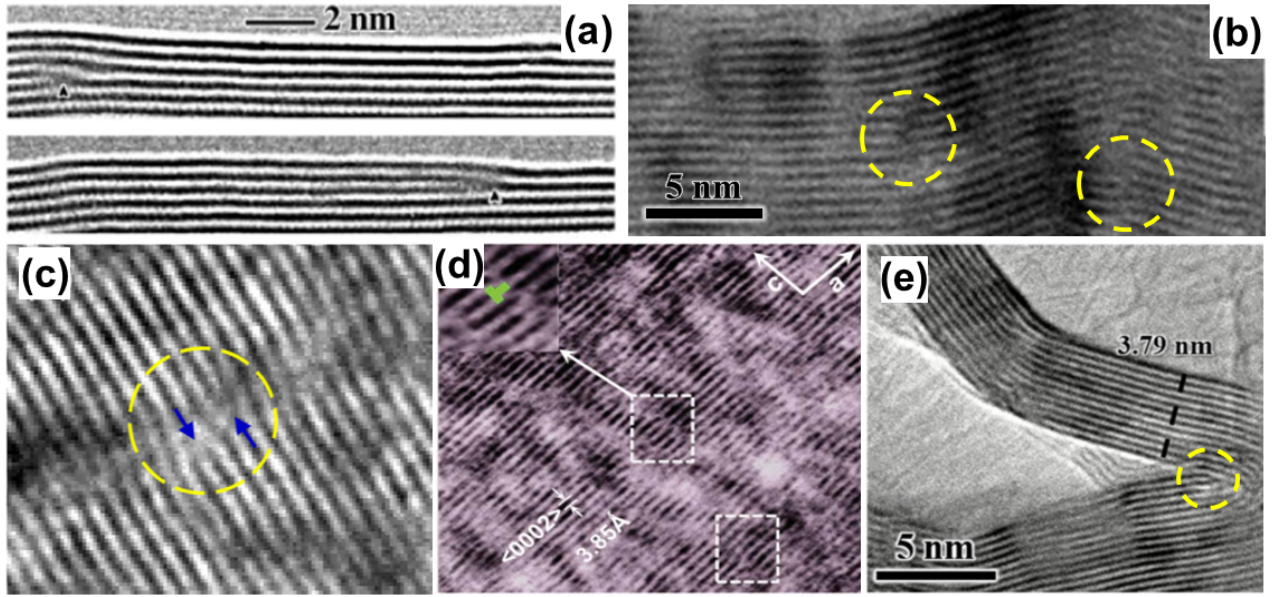

**Fig. 19:** Experimental literature HRTEM images showing: (a) two snapshots of prismatic dislocation glide in a carbon nanotube. Reprinted from Ref. [16], Copyright (2007), with permission from John Wiley & Sons. This structure is similar to Fig.17(a-c). (b) one region with fully $sp^3$ inter-core bond. Reprinted from Ref. [56], Copyright (2020), with permission from Elsevier. This structure is similar to the metastable structure shown in Fig.15(d). (c) The co-existence of both ILB and FSE configurations. Reprinted from Ref. [14], Copyright (2011), with permission from Elsevier. The modelling of these structures are depicted in Fig.15(a). (d) The presence of $sp^3$ zone with dislocation climb within prismatic dislocation complexes (image reproduced from [57]). These structures match the modelled counterparts illustrated in Fig.15(d) and Fig.17(b), respectively. (e) A region of large c-axis tilt in irradiated graphite, accompanied by material which is visibly similar to the bonded prismatic dislocation core. Reprinted from Ref. [18], Copyright (2019), with permission from Elsevier.

## 4. Discussion and Conclusions

In this work, we have performed extensive ab-initio calculations of prismatic dislocated cores, in isolated and bulk supercells. There are a number of implications of the calculations which are potentially relevant to models of graphite deformation subject to irradiation and other damage sources. For both isolated and bulk cells, the lowest energy structure lies along the armchair direction, which agrees well with previous theoretical and experimental results for prismatic loops, graphite step edges and graphene nanoribbons edges [21, 38, 44]. The lowest energy armchair core structure involves the formation of interlayer bonds in order to eliminate dangling bonds, while for the zigzag core we find low energy free-standing Dienes reconstructed cores, and zig-zag interlayer bonded cores. Along both crystallographic orientations, differences in line energy are relatively small, and it is likely that multiple edge configurations can occur.

Local stacking conditions influence the prismatic core structure. Our calculations of isolated cores find higher energies for all AB-stacked prismatic cores due to strain and disruption of perfect aromaticity within the core. The small line energy of a basal dislocation in graphite, of the order of 0.07 eV/Å, is substantially less than the energy differences between AA- and AB-stacked prismatic cores [45], suggesting that AA-stacked cores with associated basal dislocations to restore AB stacking in the lattice are most likely to be the dominant reconstruction. In the zigzag case the free-standing edge is stabilised by bond rotations along the core, which cannot be transferred during dislocation motion and will effectively freeze dislocation glide until the zigzag edge is restored.

Bulk calculations of prismatic cores were performed in quadrupole and grain boundary configurations. Importantly, we find that while high-angle grain boundaries result in stable, low-energy prismatic dislocation arrays, at low angles the quadrupolar configuration is preferred, suggesting that regular grain boundary arrays are less likely at smaller inter-grain angles. Notably, at intermediate angles we find bonding configurations with local double bonds which are extremely similar to the reconstructed Pandey diamond surfaces, occurring within the bulk of rotationally disordered graphite.

Our NEB calculations provide significant insight into the microscopic processes underlying different material deformation modes. Calculated energy barriers indicate that interlayer bonds are highly stable, and significant bond breaking can only occur at high temperatures. This barrier also sets constraints on the motion of the prismatic dislocation. Prismatic dislocations and grain boundaries of c-axis character are therefore likely to be highly immobile until high temperatures ($T > 2000K$), a result which agrees well with experimental results for prismatic dislocations in carbon nanotubes. [16]

The observation that a prismatic core can climb via *interlayer sliding*, rather than (or in addition to) the classical vacancy and interstitial trap-and-release mechanism seen in isotropic 3D materials may have important implications for irradiation-induced changes. In principle, prismatic climb, by the sliding of small loops against adjacent layers, could lead to the generation of basal dislocations in the vicinity of prismatic loops in bulk. This suggests a possible mechanism which could lead to the generation of basal dislocations, explaining the star-shaped patterns seen on the surface of irradiated HOPG at low temperatures [58], and which can have a profound effect on irradiation-induced volumetric change. [55, 59]

Finally, our calculations of the barrier to anneal a pair of prismatic cores is essentially the same as the barrier to perform glide, and we conclude that these defects are unlikely to anneal out until the same high temperatures where prismatic glide can occur. While we find very little additional energy is required in order for pairs of oppositely-signed cores to annihilate, we do find metastable structures which can impose substantial energy barriers toward the realignment and annealing in the core region. These more disordered $sp_3$-bonded regions are a likely contributor to the unannealable damage evident in graphite irradiated at high temperatures. [53] The energy release involved when these anneal out could be significant, and may also be a contributing factor to Wigner energy.

There are a number of interesting future research directions on the basis of this work. The biggest limitation of our ab-initio calculations is the cell size. Currently, the most commonly used graphene potentials do not correctly capture the $sp^3$ bonding, and consequently cannot model the prismatic dislocation core [60]. This severely inhibits our ability to examine certain properties, which in principle would require multiple simulations with, minimally, thousands of carbon atoms to accurately map out the entire energy range. Future work which could both accurately and more efficiently model the chemical interactions of graphene edges, for example through the development and use of more accurate molecular dynamics potentials, would do much to illuminate issues related to the diffusion and kinking of dislocation cores.

This would also open the door to examine a wider variety of dislocation structures, with intermediate combinations of dislocation line and Burgers vector, and would allow the direct simulation of prismatic loops. There are a number of other interesting questions suggested by our work, including questions regarding the full range of grain boundary structures at intermediate angles, the interaction of environmental impurities with dislocation cores, kinks, jogs and diffusion, and the magnetic and electronic properties of prismatic cores, which have the potential to be similar to the novel states observed at graphene nanoribbon edges. The study of the prismatic dislocation is overall of immense foundational interest, is likely immediately relevant to models of graphite damage, and is deserving of significant further work.

**Acknowledgements**

This work was supported by the United Kingdom EPSRC grant EP/R005745/1, Mechanisms of Retention and Transport of Fission Products in Virgin and Irradiated Nuclear Graphite. Kenny Jolley and Pavlos Mouratidis also gratefully acknowledge funds from EDF energy generation 2016-2021. The authors gratefully acknowledge the use of Athena at HPC Midlands+, which was funded by the EPSRC grant EP/P020232/1 as part of the HPC Midlands+ consortium. CE and AI acknowledge ANR-16-CE24-0008-01 "EdgeFiller" and ANR-20-CE08-0026 "OPIFCat" for funding. DE acknowledges support from the TUBITAK-2219 post-doctoral research abroad fund.

# SUPPLEMENTARY MATERIALS

# Prismatic Edge Dislocations in Graphite

James G. McHugh[1,2,3*], Pavlos Mouratidis[1], Anthony Impellizzeri[4],
Kenny Jolley[1], Dogan Erbahar[4,5*], Chris P. Ewels[4]

[1]*Department of Chemistry, Loughborough University, Loughborough, LE11 3TU, UK*

[2]*Department of Physics and Astronomy, University of Manchester, Manchester, UK,*

[3]*National Graphene Institute, University of Manchester, Manchester, UK.*

[4]*Université de Nantes, CNRS, Institut des Matériaux Jean Rouxel, IMN, F-44000 Nantes, France*

[5]*Dogus University, Faculty of Engineering, Department of Mechanical Engineering, Ümraniye, 34775, İstanbul, Turkey*

## 1. Isolated RK Edges

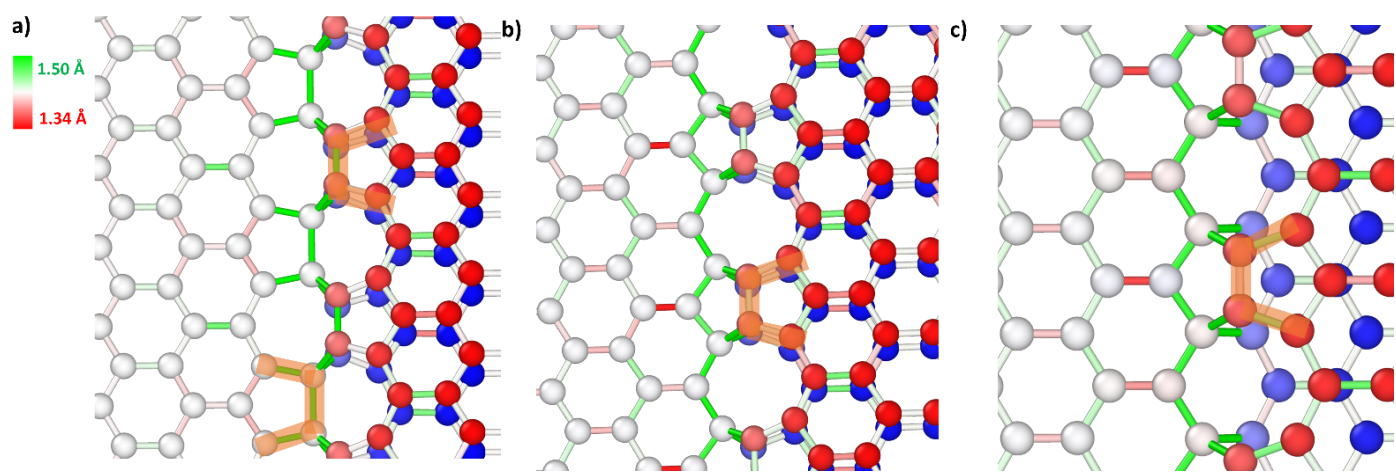

**Supplementary Figure 1**: **a)** RK-RK-RK core, where monolayer (ML) and both bilayer (Top, Bottom) regions have reconstructed Klein edges. **b)** ZZ-RK-RK core and **c)** ZZ-ZZ-RK cores, which result in AB-stacking. C-C bonds are coloured depending on the degree of bond strain as shown in the key on the left, with white matching the pristine graphene bond length (1.46 Å). Reconstructed pentagons are indicated as a guide for the eye.

## 2. Quadrupole & Monoclinic Boundary Conditions

Having considered the relative energies of different prismatic cores for the same unit cell, we briefly examine the dependence of the calculated energies on cell dimensions for the two low energy bonded configurations (ZZ-ZZ-ZZ and AC-AC-AC) in the quadrupole configuration, using different orthorhombic and monoclinic boundary conditions. These energies are summarised in **Supplementary Table 1 (Fig. 9g** defines the structural parameters $L_1$ and $L_2$). There is moderate variation in the predicted core line energy between different cell configurations due to varying elastic energy, but we generally observe consistent formation line energies within a range of 0.1 eV/Å of one another, with a markedly smaller variation for the armchair cores.

| Direction | Cell | $L_1$ (nm) | $L_2$ (nm) | $E_f$ (eV/Å) |
|---|---|---|---|---|
| **Zigzag** | Orthorhombic | 1.990 | 1.249 | 1.590 |
|  | Monoclinic | 2.130 | 1.229 | 1.429 |
|  | Monoclinic | 3.133 | 1.240 | 1.490 |
|  | Monoclinic | 3.132 | 1.580 | 1.577 |
| **Armchair** | Orthorhombic | 1.120 | 1.507 | 1.294 |
|  | Monoclinic | 1.120 | 1.507 | 1.296 |
|  | Monoclinic | 1.300 | 1.507 | 1.292 |

**Supplementary Table 1:** Formation line energies, $E_F$, (eV/Å), for the interlayer bonded armchair and zigzag cores in different cell configurations in bulk graphite. $L_1$ and $L_2$ refer to the dipole quadrupole spacings (see Fig. 9e). **Δ$E_f$** refers to the energy difference between these interlayer bonded structures and equivalent non-bonded structures within the crystal (positive values indicate the ILB structure to be more stable).

## 3. Angular Variation of Grain Boundary Structures

We include here some additional images of simulated grain boundary structures.

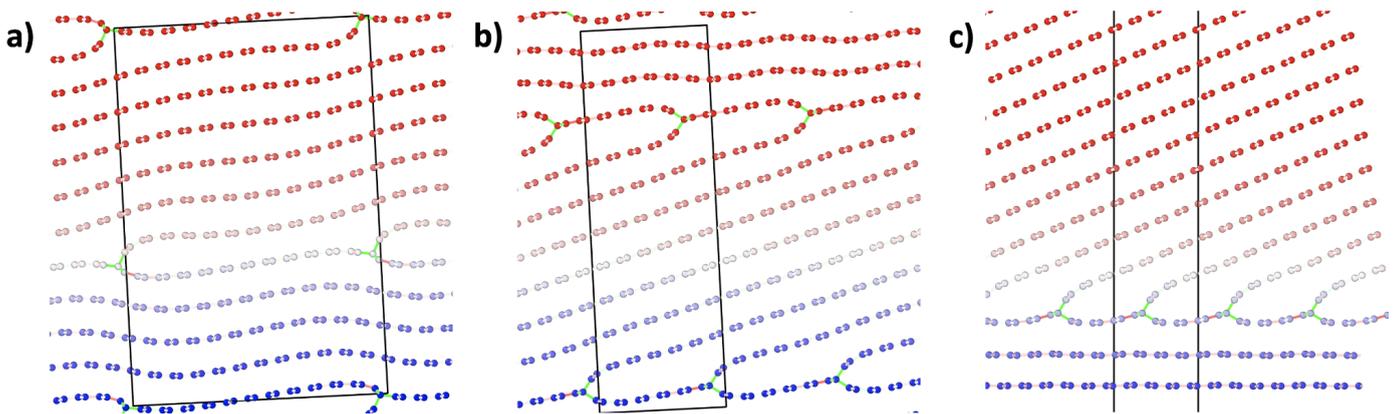

**Supplementary Figure 2:** Zigzag grain boundaries with **a)** θ = 7.8°, **b)** θ = 15.8° and **c)** θ = 25°. Notably there is substantially more rippling around the cores as the angle decreases due to inter-gran strain at the boundary.

## 4. Pandey-reconstructed Grain Boundary Surfaces

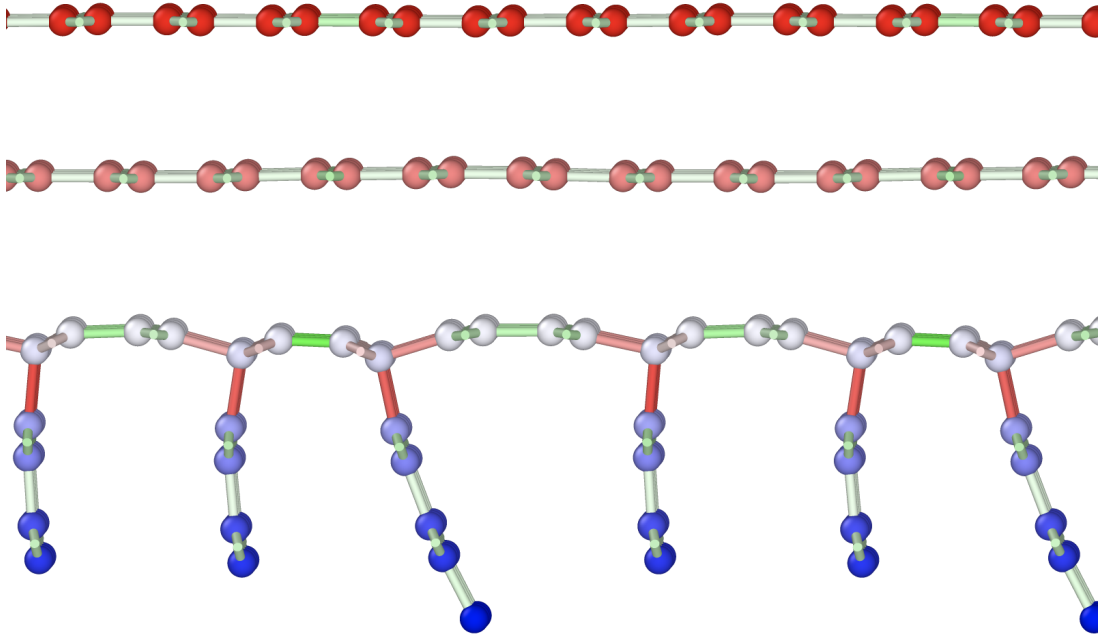

**Supplementary Figure 3:** 60° zigzag grain boundary. Central C-C pairs connected via short (green) bonds are reminiscent of Pandey-π bonded chains, consisting of isolated pairs of double bonded carbon atoms.

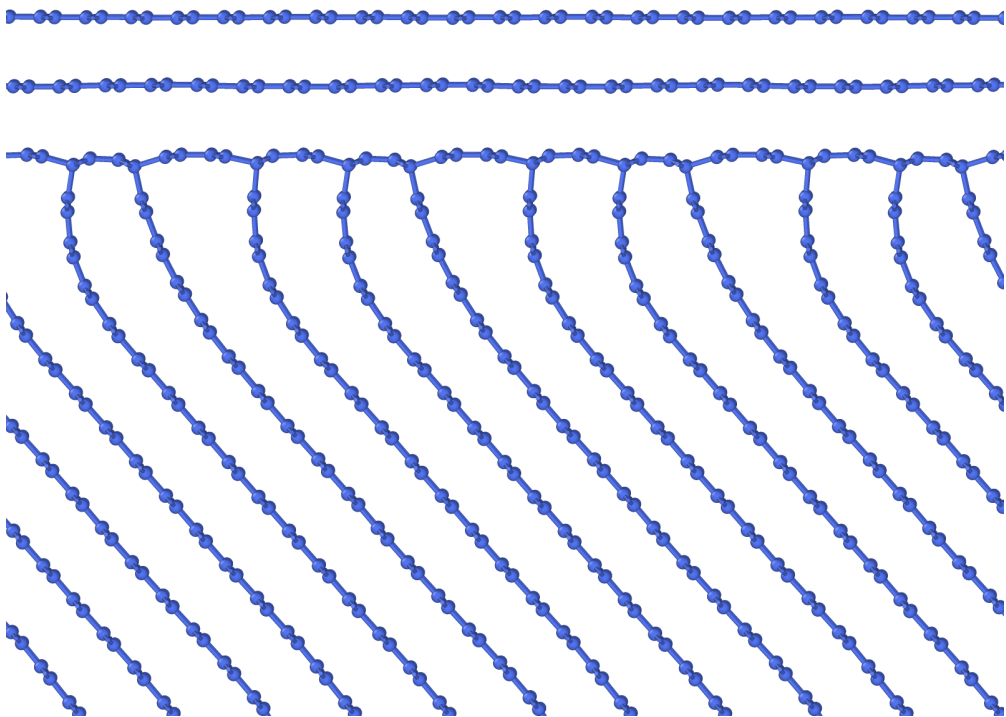

**Supplementary Figure 4:** 60° zigzag grain boundary

## 5. Bond-breaking probability

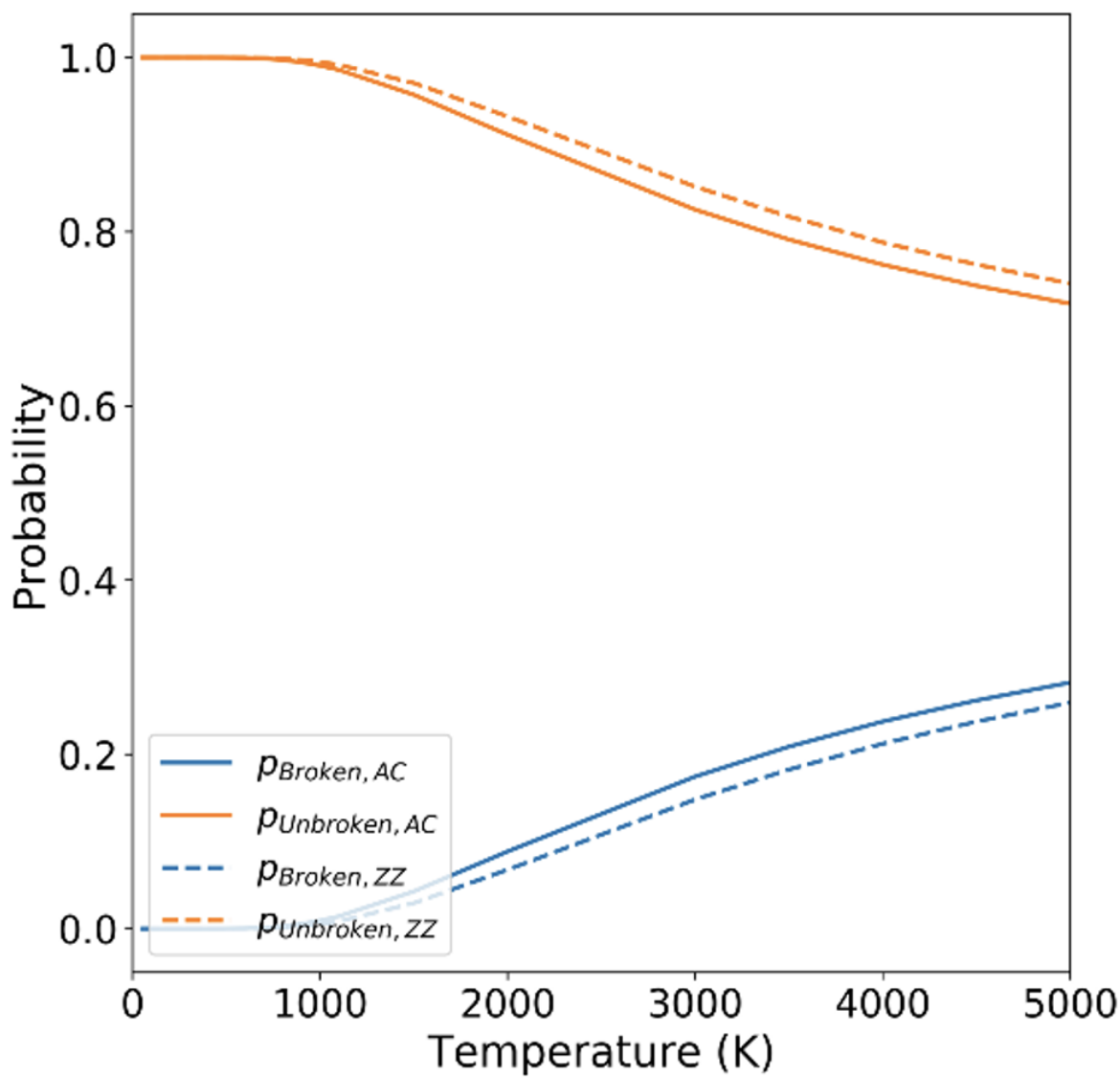

**Supplementary Figure 5:** Bond-breaking probability vs temperature, calculated using Equation (2) and CI-NEB barriers for the armchair and zigzag cores.

## 6. Isolated, extended dislocation supercells

To examine the dynamics of the bond-breaking process, we have created supercells which are long in the direction of the dislocation line. These cells are created using the initial isolated core structures, which have been repeated along the dislocation line (y) direction. Even for isolated cells this results in very large supercells when a full dislocation dipole is considered. To circumvent this, we have created a smaller cell containing only one dislocation core by cutting out a segment of the full isolated cell, freezing carbon atoms far from the core in their equilibrium configuration, and passivating the resulting dangling bonds. This is shown in **Fig. S6**.

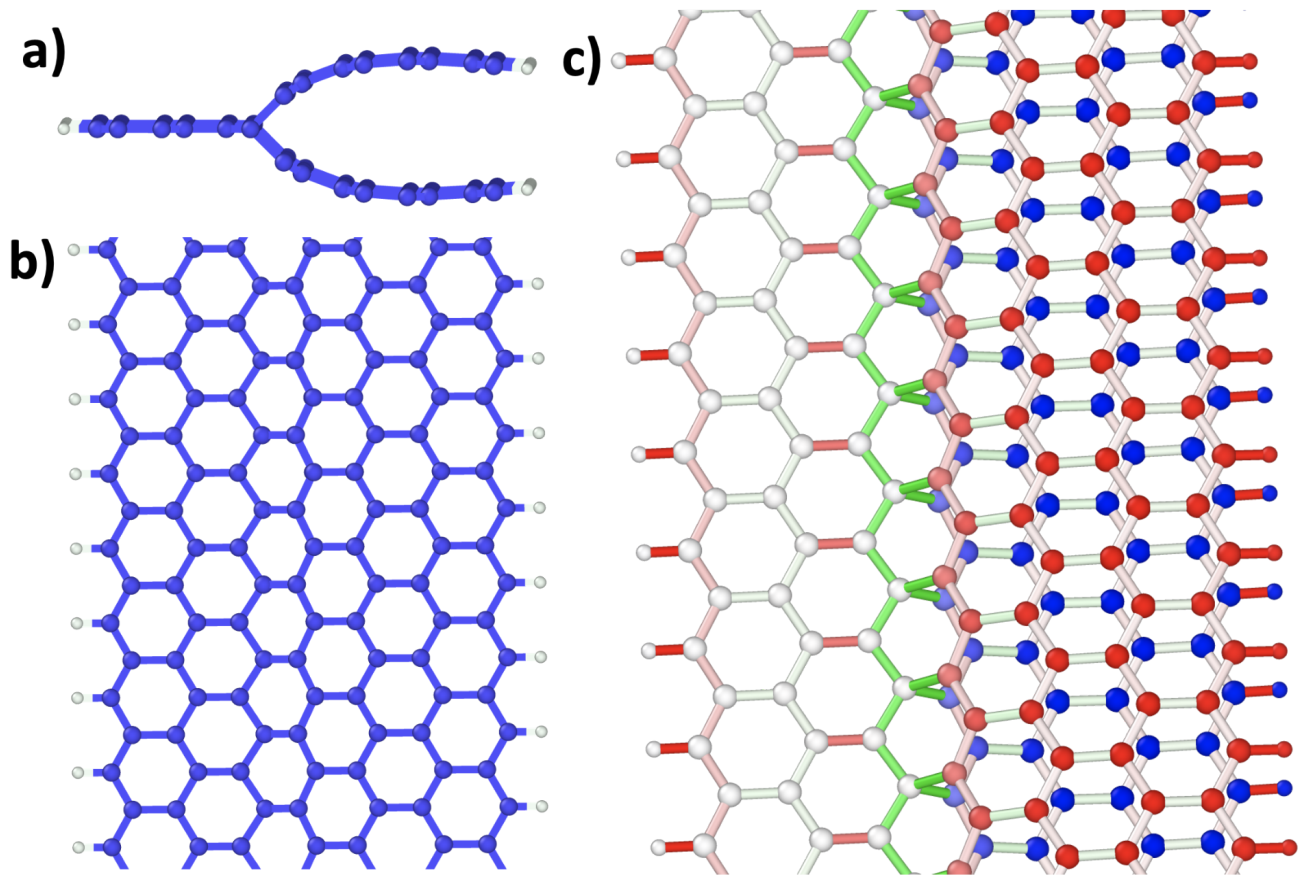

**Supplementary Figure 6:** Isolated cell used to simulate bond breaking along dislocation core. **a)** Top view **b)** Side view **c)** Final, intermediate view of relaxed, unbroken cell.

# 7. Defocus analysis of simulated TEM images

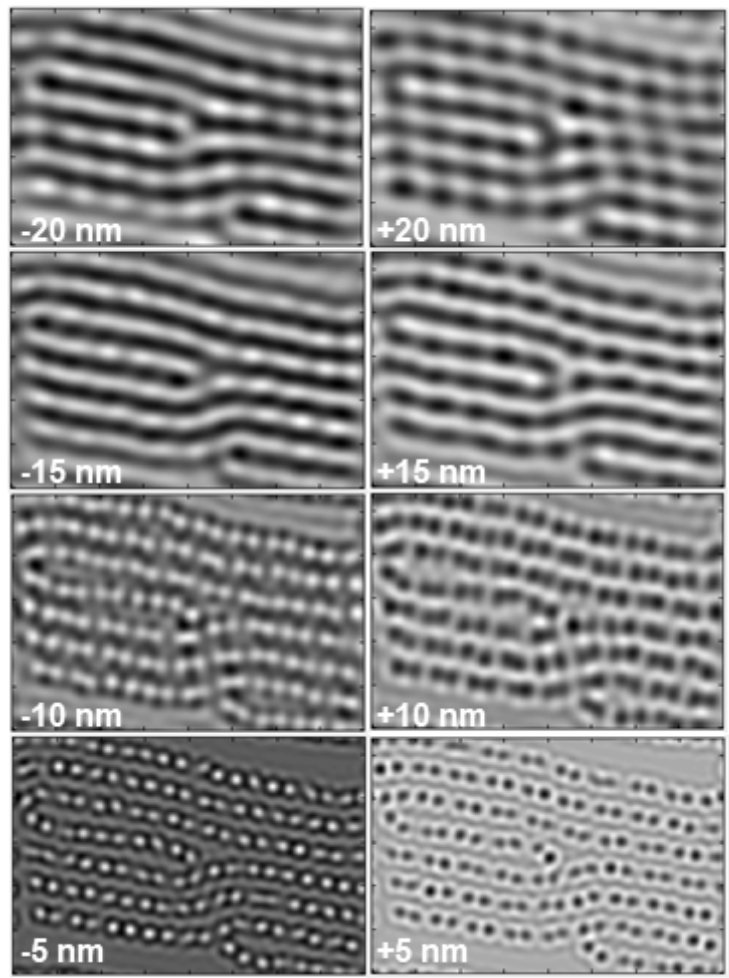

**Supplementary Figure 7:** TEM simulated images of bonded prismatic core dislocation as a function of the defocus.

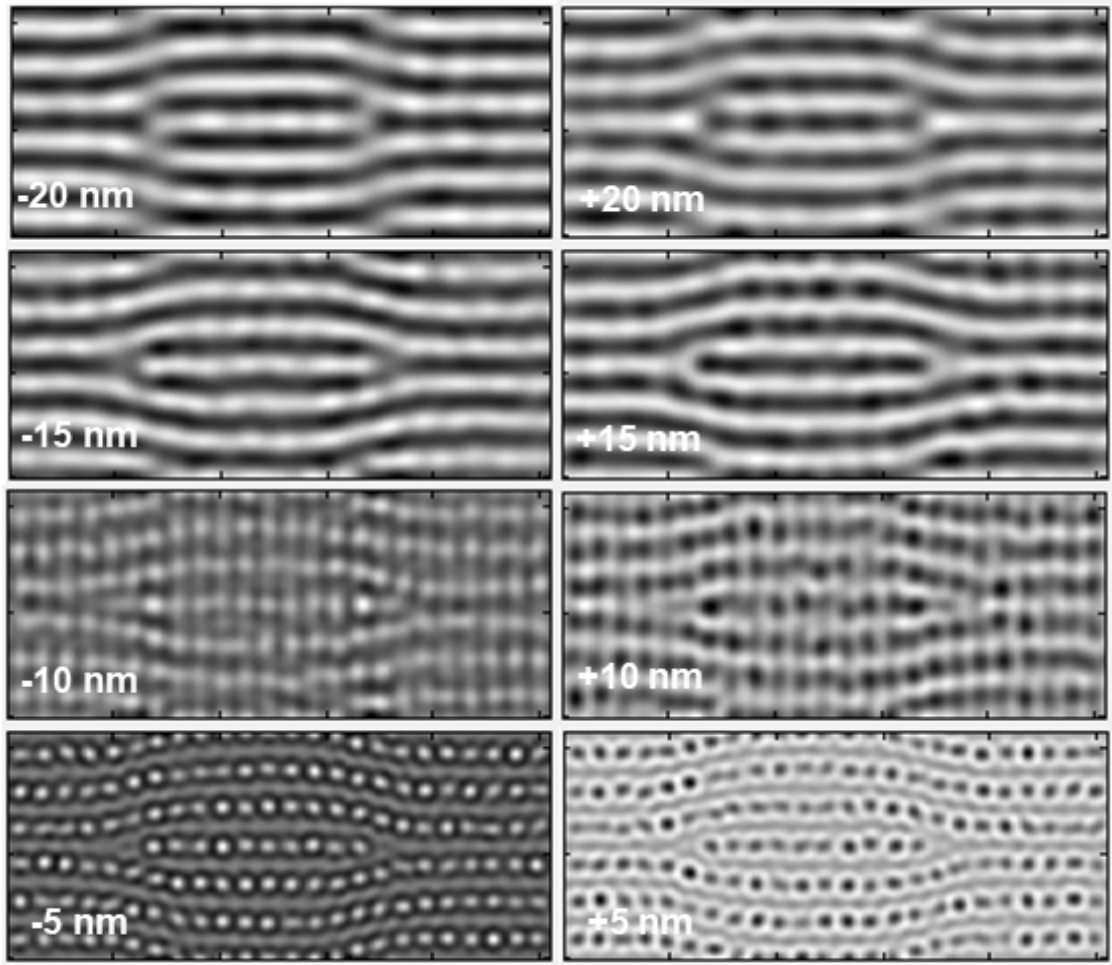

**Supplementary Figure 8:** TEM simulated images of broken prismatic core dislocation as a function of the defocus.

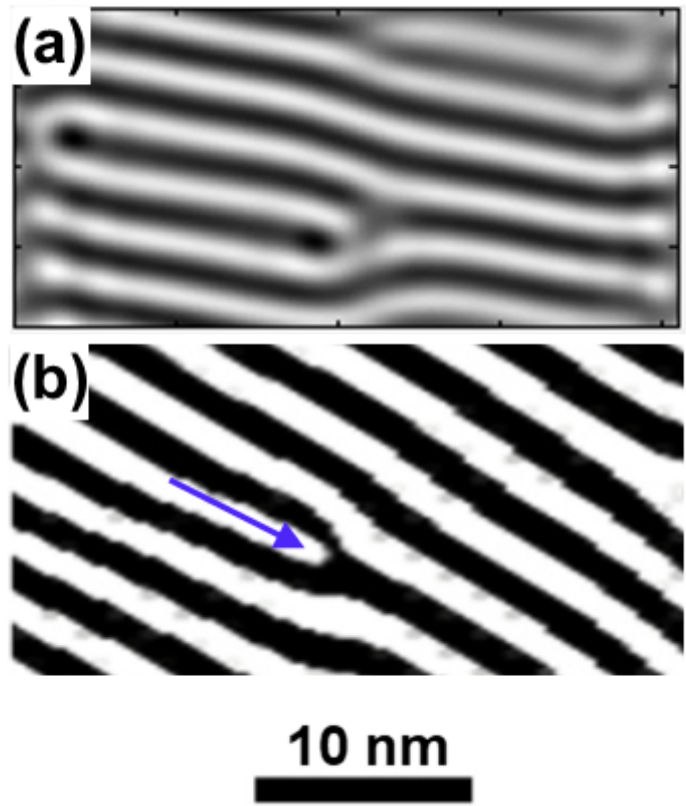

**Supplementary Figure 9:** (a) Simulated HRTEM image of bonded prismatic core dislocation without thermo-diffuse scattering. (b) Experimental HRTEM image of an equivalent structure reprinted from Karthik, C., Kane, J., Butt, D. P., Windes, W. E., and Ubic, R. (2011). In situ transmission electron microscopy of electron-beam induced damage process in nuclear grade graphite. Journal of Nuclear Materials, 412(3), 321-326, Copyright (2011), with permission from Elsevier.

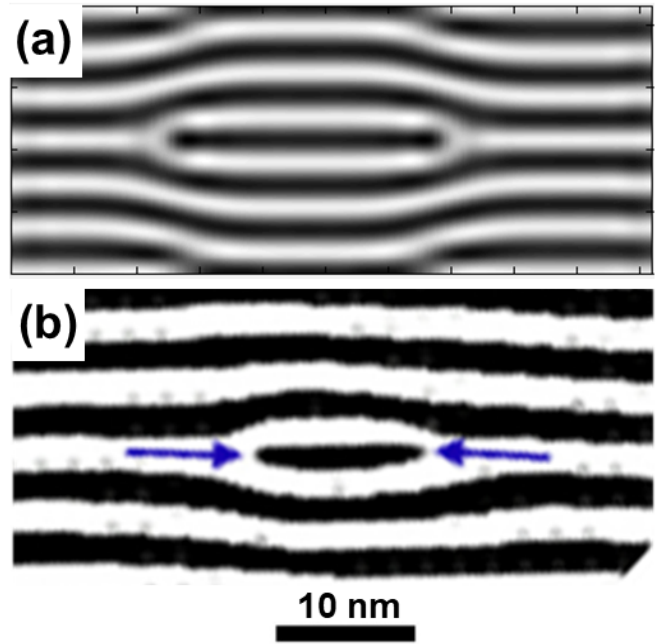

**Supplementary Figure 10:** (a) Simulated HRTEM image of broken prismatic core dislocation without thermo-diffuse scattering. (b) Experimental HRTEM image of an equivalent structure reprinted from Karthik, C., Kane, J., Butt, D. P., Windes, W. E., and Ubic, R. (2011). In situ transmission electron microscopy of electron-beam induced damage process in nuclear grade graphite. Journal of Nuclear Materials, 412(3), 321-326, Copyright (2011), with permission from Elsevier.